\documentclass[a4paper,11pt]{article}
\usepackage{aaskaiid}
\usepackage{orcidlink}

\newcommand{\hi}{\textrm{H\textsc{i}}}
\newcommand{\Ohi}{\Omega_{\textrm{H\textsc{i}}}}
\newcommand{\bhi}{b_{\textrm{H\textsc{i}}}}
\newcommand{\secref}[1]{\hyperref[#1]{Section~\ref*{#1}}}
\newcommand{\appref}[1]{\hyperref[#1]{Appendix~\ref*{#1}}}
\newcommand{\txteq}[1]{\,{#1}\,}

\newcommand{\degsq}{\,\deg^2}
\title{Single-dish \hi\ Intensity Mapping with the SKAO: Precursor Progress with MeerKAT's Large Area Synoptic Survey (MeerKLASS)}
\ShortTitle{Single-dish IM with the SKAO}

\author[1,2]{Steven Cunnington\orcidlink{0000-0001-6594-107X}}
\ShortName{Cunnington \& Wang et al.} 
\author[3,4]{Jingying Wang\orcidlink{0000-0002-5598-2668}}
\author[4,5]{Mario G. Santos\orcidlink{0000-0003-3892-3073}}
\author[6,7,8,9]{Matilde Barberi-Squarotti\orcidlink{0009-0007-8964-5807}}
\author[2,4]{Philip Bull}
\author[9,10,11,4]{Stefano Camera\orcidlink{0000-0003-3399-3574}}
\author[12,13]{Isabella P. Carucci\orcidlink{0000-0001-5287-0065}}
\author[14]{Zhaoting Chen\orcidlink{0000-0002-4965-8239}}
\author[4]{Brandon Engelbrecht}
\author[4,15,16]{José Fonseca}
\author[4]{Karin Fornazier\orcidlink{0000-0003-0578-9533}}
\author[2]{Keith Grainge}
\author[4]{Wenkai Hu\orcidlink{0000-0002-3108-5591}}
\author[17,4]{Melis O. Irfan\orcidlink{0000-0003-2021-7357}}
\author[18,4]{Yichao Li\orcidlink{0000-0003-1962-2013}}
\author[14,4]{Alkistis Pourtsidou\orcidlink{0000-0001-9110-5550}}
\author[19,4]{Marta Spinelli}
\author[4,2]{Amadeus Witzemann}
\author[2]{Laura Wolz}

\affiliation[1]{Institute of Cosmology \& Gravitation, University of Portsmouth, Dennis Sciama Building, Portsmouth, PO1 3FX, UK}
\affiliation[2]{Jodrell Bank Centre for Astrophysics, Department of Physics \& Astronomy, The University of Manchester, Manchester M13 9PL, UK}
\affiliation[3]{Shanghai Astronomical Observatory, Chinese Academy of Sciences, 80 Nandan Road, Shanghai 200030, China}
\affiliation[4]{Department of Physics \& Astronomy, University of the Western Cape, Cape Town 7535, South Africa}
\affiliation[5]{South African Radio Astronomy Observatory (SARAO), Cape Town, 7700, South Africa}
\affiliation[6]{Dipartimento di Fisica, Universit\`a degli Studi di Milano, via G.\ Celoria 16, 20133 Milano, Italy}
\affiliation[7]{INFN – Istituto Nazionale di Fisica Nucleare, Sezione di Milano, via G.\ Celoria 16, 20133 Milano, Italy}
\affiliation[8]{INAF -- Istituto Nazionale di Astrofisica, Osservatorio Astrofisico di Brera-Merate, via Brera 28, 20121 Milano, Italy}
\affiliation[9]{Dipartimento di Fisica, Universit\`a degli Studi di Torino, via P.\ Giuria 1, 10125 Torino, Italy}
\affiliation[10]{INFN – Istituto Nazionale di Fisica Nucleare, Sezione di Torino, via P.\ Giuria 1, 10125 Torino, Italy}
\affiliation[11]{INAF – Istituto Nazionale di Astrofisica, Osservatorio Astrofisico di Torino, 10025 Pino
Torinese, Italy}
\affiliation[12]{INAF -- Istituto Nazionale di Astrofisica, Osservatorio Astronomico di Trieste, Via G.B.\ Tiepolo 11, 34131 Trieste, Italy}
\affiliation[13]{IFPU -- Institute for Fundamental Physics of the Universe, Via Beirut 2, 34151 Trieste, Italy}
\affiliation[14]{Institute for Astronomy, The University of Edinburgh, Royal Observatory, Edinburgh EH9 3HJ, UK}
\affiliation[15]{Departamento de F\'isica e Astronomia, Faculdade de Ci\^{e}ncias, Universidade do Porto, CAUP, Rua do Campo Alegre 687, Porto, 4169-007, Portugal}
\affiliation[16]{Instituto de Astrof\'isica e Ci\^encias do Espa\c{c}o, Universidade do Porto CAUP, 4150-762 Porto, Portugal}
\affiliation[17]{Institute of Astrophysics, University of Cambridge, Madingley Road, CB3 0HA, UK}
\affiliation[18]{Department of Physics, College of Sciences, Northeastern University, Wenhua Road, Shenyang 11089, China}
\affiliation[19]{Observatoire de la Côte d’Azur, Laboratoire Lagrange, Bd de l’Observatoire, CS 34229, 06304 Nice cedex 4, France}
\emailAdd{steve.cunnington@port.ac.uk}
\emailAdd{jywang@shao.ac.cn}

\abstract{Using the SKAO to map the intensity of neutral hydrogen’s 21\,cm emission line will be a golden opportunity to constrain models of cosmology. To access the largest cosmological scales, wide-sky surveys should ideally reach thousands of square degrees, requiring SKA-Mid's dishes to scan the sky in auto-correlation mode, so-called \textit{single-dish} observations. In this chapter, we overview the latest results from MeerKAT's Large Area Synoptic Survey (MeerKLASS), which has been pioneering this single-dish observing strategy, and motivating its continuation with the SKA-Mid AA4 deployment. MeerKLASS, operating on the same Karoo site where the SKA-Mid is being built, has now achieved multiple cosmological detections from single-dish observations, including high-significance cross-correlations with optical galaxy surveys and continually improving measurements of the \hi\ auto-power spectrum. These results demonstrate that stable calibration, effective foreground mitigation, and statistical recovery of cosmological signal are all achievable with a large multi-dish telescope in total-power mode. The success of MeerKLASS therefore validates the observational strategies required for SKA-Mid and marks a key milestone in demonstrating the viability of single-dish \hi\ intensity mapping for cosmology. Looking ahead, SKA-Mid’s increased sensitivity and Band 1 coverage (350–1050 MHz) will allow the same methodology to probe redshifts up to $z\txteq{\sim}3$, mapping volumes several orders of magnitude larger than currently accessible. The techniques refined with MeerKLASS thus form the operational and scientific foundation for a large portion of the SKAO’s cosmology programme.}

\begin{document}
\maketitle

\section{Introduction}

The largest cosmological scales in our Universe are among the most promising sources of information for testing the fundamental physics of cosmology. On these vast scales, the density field remains in the linear regime, largely free from the astrophysical complexities of non-linear structure formation that infiltrate smaller scales. These modes therefore preserve the cleanest imprints of cosmological physics, offering opportunities to probe phenomena such as baryonic acoustic oscillations (BAO) \citep{Eisenstein:1997ik}, redshift space distortions (RSD) \citep{Scoccimarro:2004tg}, modified gravity \citep{Koyama:2015vza}, primordial non-Gaussianity \citep{Dalal:2007cu} and relativistic effects \citep{Bonvin:2011bg}. Unlocking this information demands wide-field surveys that can observe coherent fluctuations over thousands of megaparsecs.

The SKA Observatory (SKAO) is uniquely positioned to access these scales through 21\,cm intensity mapping of neutral hydrogen (\hi) \cite{Wolz01.2026.SKA,Fonseca01.2026.SKA}. Rather than resolving individual galaxies, this technique measures the integrated emission of the 21\,cm line across broad regions of sky and frequency, enabling an efficient three-dimensional mapping of large-scale structure across vast cosmic volumes \citep{Bharadwaj:2000av,Battye:2004re,Wyithe:2007rq,Chang:2007xk,Liu:2019awk,Villaescusa-Navarro:2018vsg}. By surveying the integrated \hi\ signal, intensity mapping can rapidly build up a tomographic view of matter distribution from the late-Universe to redshifts that, in principle, can extend into the epoch of reionisation and even the dark ages \citep{Furlanetto:2006jb,Zaroubi:2012in,Kovetz:2017agg}.

\textcolor{black}{The feasibility of 21cm intensity mapping has already been demonstrated using pre‑existing facilities, with early measurements from the Green Bank Telescope \citep{Masui:2012zc,Switzer:2013ewa,Wolz:2015lwa,eBOSS:2021ebm} and Parkes \citep{Anderson:2017ert} providing the first detections of large‑scale \hi\ fluctuations. Building on this foundation, several experiments have since been commissioned with 21cm intensity mapping as a science goal. MeerKAT \citep{2016mks..confE...1J}, SKAO's precursor and the focus of this chapter, was completed in 2018 and has now delivered multiple cosmological results using 21cm intensity mapping \citep{2021MNRAS.505.3698W}, which we review throughout this work. These include high‑significance cross‑power spectrum detections with overlapping galaxy surveys \citep{Cunnington:2022uzo,mPCA,2025MNRAS.537.3632M}, stacked \hi\ emission detections \citep{2025MNRAS.537.3632M,2025arXiv250403908C}, and a measurement of the \hi\ auto‑correlation signal on sub‑Mpc scales \citep{Paul:2023yrr}. Similarly, CHIME \citep{CHIME:2022dwe}, which also began science operations in 2018, has reported stacked \hi\ emission detections in galaxy cross‑correlation \citep{CHIME:2022kvg,CHIME:2023til} and has also recently presented auto‑correlation measurements \citep{CHIME:2025cee}, again primarily on smaller‑scale modes. Alongside these, other ongoing efforts with FAST \citep{2020MNRAS.493.5854H,Li:2023zer} and uGMRT \citep{Elahi:2024qbp} are steadily progressing toward precision cosmological analyses, collectively demonstrating the rapid maturation of 21cm intensity mapping across a diverse set of instruments.} We refer to \cite{Elahi01.2026.SKA} for a wider discussion on other SKAO pathfinders.

\textcolor{black}{Accessing ultra-large scales in 21cm intensity mapping} with the SKA-Mid \textcolor{black}{will require} operating the dishes in \textit{auto-correlation} (or single-dish) mode \citep{2015aska.confE..19S}. The interferometer baselines of SKA-Mid are optimised for high-resolution imaging, but lack sensitivity to the largest angular scales corresponding to features such as the BAO scale at ${\sim}\,100\,$Mpc \citep{Bull:2014rha}. Single-dish observations, on the other hand, preserve the wide fields of view and continuous spatial information needed for intensity mapping at these scales. This observing mode, combined with efficient sky-scanning, offers a powerful pathway to measure the clustering of \hi\ over tens of thousands of square degrees.

Fortunately, the SKAO’s South African-based precursor, MeerKAT\footnote{\href{https://www.sarao.ac.za/science/meerkat/}{sarao.ac.za/science/meerkat}}, is already demonstrating the success of this approach. MeerKAT, operated by the South African Radio Astronomy Observatory (SARAO), is located on the same site as the future SKA-Mid array in the Karoo desert and will eventually be incorporated into it around the turn of the decade. Through MeerKAT’s Large Area Synoptic Survey (MeerKLASS)\footnote{\href{https://meerklass.org/}{meerklass.org}} \citep{MeerKLASS:2017vgf}, the telescope has been pioneering single-dish \hi\ intensity mapping, delivering both technical validation and early cosmological results \citep[see][for a recent review]{Cunnington:2025sdr}, while at the same time delivering a wide, high-resolution continuum survey using the interferometer, as discussed in \citet{Chatterjee01.2026.SKA} \citep[also see][]{Chatterjee:2025wzm,Paul:2025cjp,Mangla:2025ans}. MeerKLASS has developed a robust calibration and analysis pipeline capable of handling complex instrumental effects, performing blind foreground cleaning, and statistically detecting the cosmological \hi\ signal. With 2,500 observing hours awarded and a target coverage of $10{,}000\degsq$ across $0.4\txteq{<}z\txteq{<}1.45$, MeerKLASS is on course to become the largest spectroscopic survey of large-scale structure in the Southern hemisphere in the pre-SKAO era.

The progress made by MeerKLASS is of critical importance for shaping early science-verification programmes with the SKAO. The observing strategy developed for MeerKLASS, i.e. operating the telescope in single-dish mode, is directly transferable to SKA-Mid, and does not depend on the long interferometric baselines that will only become available in later deployment phases. Because single-dish intensity mapping requires only total-power measurements from each antenna, the data rates and processing requirements are dramatically reduced compared to conventional interferometric observations, making this technique ideally suited for early operations with the SKA-Mid AA$^*$ configuration. Moreover, SKA-Mid will provide a significant upgrade in raw sensitivity relative to MeerKAT, while its Band 1 receivers (covering 350–1050 MHz) will extend the accessible redshift range to $z\txteq{\sim}3$, allowing cosmological volumes several orders of magnitude larger to be mapped, than currently analysed data sets. Demonstrating the robustness of the single-dish \hi\ intensity mapping approach with MeerKLASS therefore provides a compelling case for making this observing mode a flagship early-science verification target for the SKA-Mid, ensuring that cosmological exploitation of the telescope can begin well before its full interferometric capabilities are online.

In this chapter, we outline the progress made by MeerKLASS and discuss how its achievements pave the way for SKAO’s cosmological intensity mapping programme. \secref{sec:Calibration} describes the MeerKLASS observing strategy and calibration methods, highlighting recent results that validate single-dish operation for large-scale cosmology, and includes a discussion on challenges of foreground removal. \secref{sec:Detections} presents the cosmological detections, mostly achieved through cross-correlation with overlapping galaxy surveys. \secref{sec:Future} connects these milestones to the future prospects for SKA-Mid, outlining how the lessons from MeerKLASS guide its cosmological science goals. We then conclude in \secref{sec:Conclusion}.

\section{Calibrating and cleaning single-dish intensity mapping observations}\label{sec:Calibration}

In this section, we describe the observing and data-processing strategies developed by MeerKLASS for 21cm single-dish intensity mapping, together with the calibration and foreground-removal techniques that underpin its cosmological analyses. These developments are directly relevant to the SKA-Mid, whose dishes will employ the same total-power observing mode to achieve the wide-area surveys required for large-scale cosmology.

The calibration and scanning procedures refined by MeerKLASS, such as constant-elevation azimuth scans, frequent noise-diode injections, and iterative self-calibration, form an end-to-end framework that ensures gain stability and accurate recovery of large-scale sky structure. Because the SKA-Mid design closely follows MeerKAT’s offset-Gregorian optical configuration and receiver chain, the same methodology will transfer almost seamlessly to the SKAO era, requiring only minor adjustments to accommodate the larger number of dishes and expanded frequency coverage.

The ability to perform reliable auto-correlation measurements, construct three-dimensional sky cubes, and remove foregrounds with blind cleaning techniques represents the essential foundation for extracting cosmological information from 21cm maps. The following subsection outlines the MeerKLASS implementation of these methods, beginning with the L-band pilot surveys, which provide the first large-scale demonstration of these calibration and cleaning pipelines in practice.

\subsection{MeerKLASS L-band pilot surveys}\label{sec:calibration_Lband}


Cosmological neutral hydrogen observations with single dishes typically require a scanning strategy where the dishes are rapidly moved across the sky. In all MeerKLASS observations so far, the 64 MeerKAT antennas were set to scan in azimuth at constant elevation to minimise fluctuations of ground spill and airmass. In one scan, the dishes are moved back and forth along azimuth to produce several scan stripes, resulting in a series of Z-shaped coverage pattern due to the sky’s rotation. Two scans can be performed per night during the target field rising and setting, so that the two scans intersect and provide good coverage in their region of overlap. Before and after each scan (typically 1.5 hours), we spent several minutes tracking a nearby celestial point source to use as a bandpass and absolute flux calibrator.  Noise diodes attached to each receiver were fired periodically during the observation to provide a relative time-ordered data (TOD) calibration reference.

\begin{figure}
\centering
\includegraphics[width=.6\columnwidth]{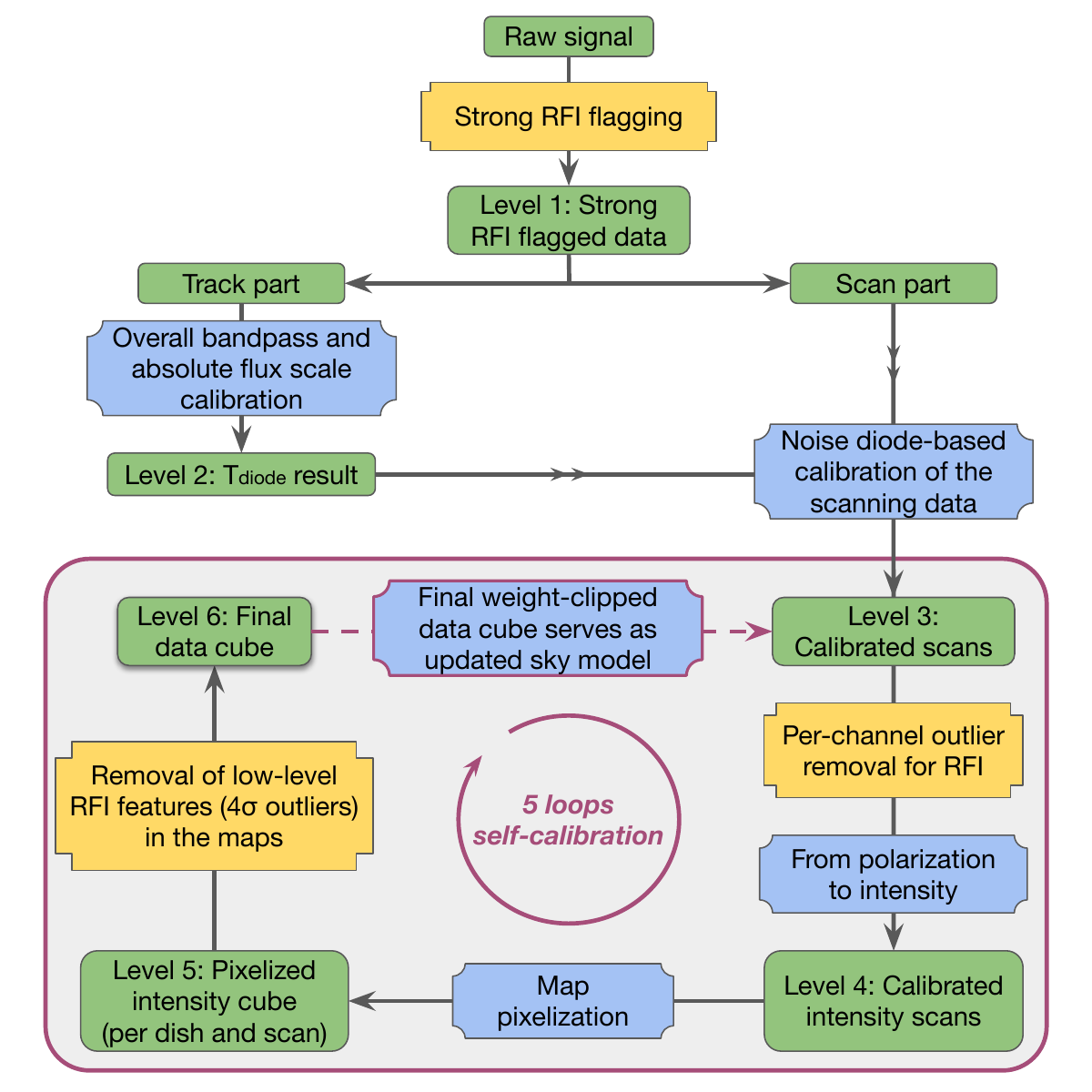}
\caption{Flowchart showing each step in the \texttt{KATcali} calibration pipeline.}
\label{fig:pipe}
\end{figure}

We build the MeerKAT single-dish calibration pipeline \texttt{KATcali}, and employ it to calibrate the MeerKLASS observations and create the combined 3D data cubes. \texttt{KATcali} includes several cycles of RFI (Radio Frequency Interference) flagging, calibration, and map-making, as shown in \autoref{fig:pipe}. 

For RFI flagging, we use the RFI package of Signal Extraction and Emission Kartographer (SEEK; \citealt{2017A&C....18....8A}), which follows the SumThreshold algorithm \citep{2010MNRAS.405..155O},  to reduce the strong RFIs, following several rounds of low-level RFI flagging in later steps (see yellow boxes in \autoref{fig:pipe}).

For calibration, we construct a model for all components that contribute to the total signal and fit the free parameters by comparing the model to the TOD using a prescribed likelihood with priors.
Our model, in temperature units, is expressed as 
\begin{gather}\label{equ:model}
  T_{\rm model}(t, \nu)= T_{\rm ps}(t,\nu)  + T_{\rm diffuse}(t, \nu) + T_{\rm el}(t, \nu)     +  T_{\rm diode}(t, \nu) 
 + T_{\rm rec} (t, \nu),
\end{gather}
where $T_{\rm ps}$, $T_{\rm diffuse}$ and $T_{\rm el}$  are the antenna temperature models of the point source contributions, celestial diffuse component (Galactic emission and the CMB), and elevation-dependent terrestrial emission (atmosphere and groundspill) respectively, while $T_{\rm diode}$ is the noise diode contribution and  $T_{\rm rec}$ is the receiver temperature.
All quantities external to the dish are considered to be already convolved by the primary beam.
To compare with the raw data  the correlator, $\hat T_{\rm raw}(t, \nu)$ where the hat indicates a temperature in the (arbitrary) correlator units, we multiply the signal model by the gain, $g(t, \nu)$, so that
\begin{gather}\label{equ:model2}
  \hat T_{\rm model}(t, \nu)= g(t, \nu)\, T_{\rm model}(t, \nu)\,.
\end{gather}
The signal and gain models are then fitted (with a \textit{Bayesian} framework; see more details in Section 3.3 of \citealt{2021MNRAS.505.3698W}) to the TOD for each polarisation (HH and VV), frequency channel, dish, and observation scan, all of which are treated independently. 
Using the gain solution, we can then obtain the calibrated temperature,
\begin{gather} \label{eq:T}
    T_{\rm cal}(t,\nu)\equiv \frac{\hat T_{\rm raw}(t, \nu)}{g(t, \nu)}\,.
\end{gather}
%

\begin{figure}
\centering
\includegraphics[width=.99\columnwidth]{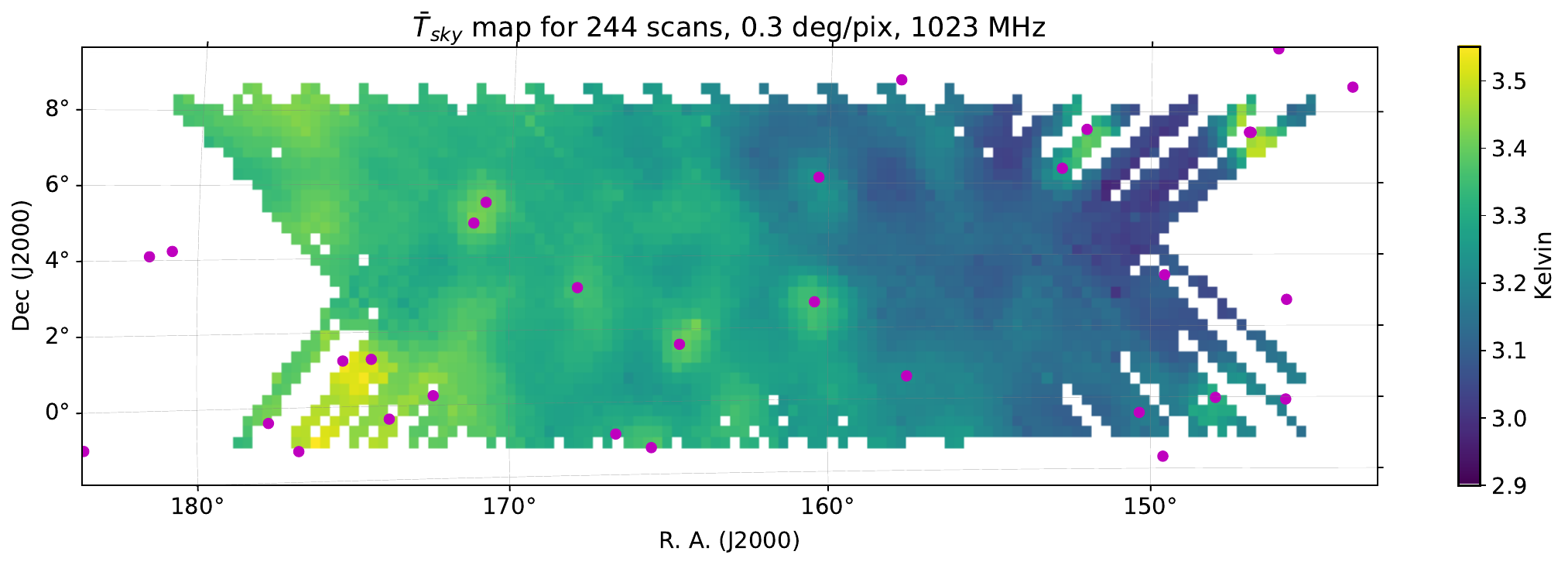}
\includegraphics[width=.99\columnwidth]{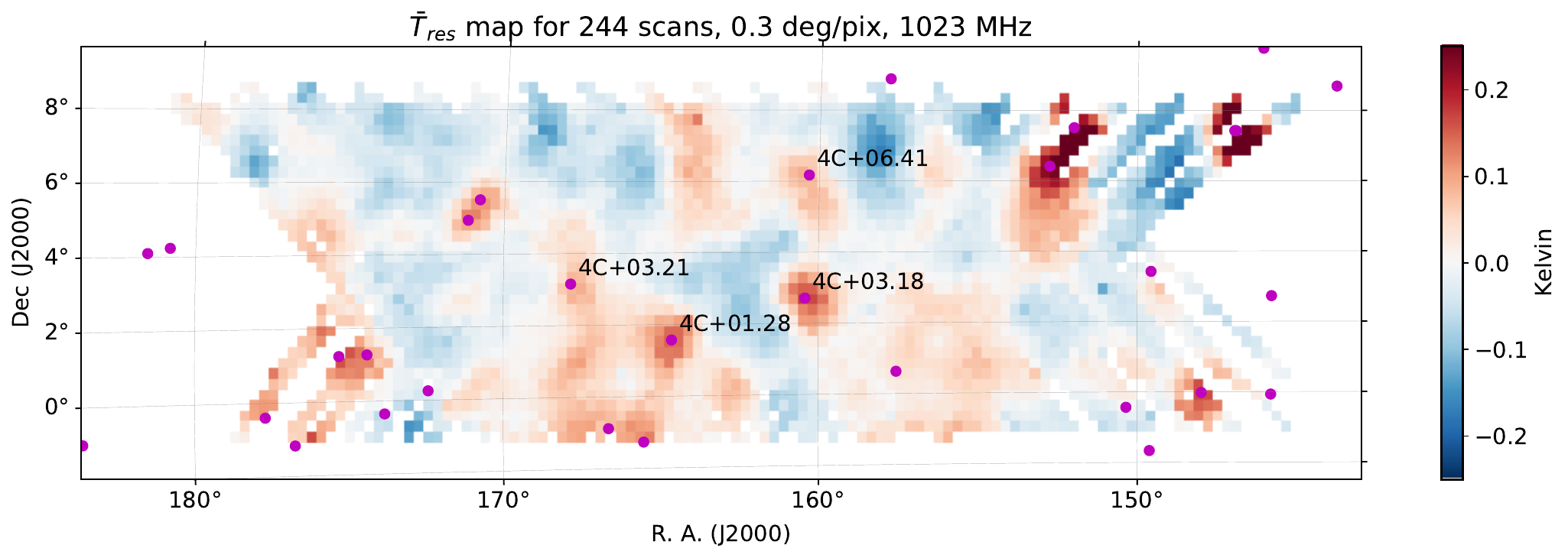}
\caption{The mean sky map ($T_{\rm sky}$) and corresponding residuals ($T_{\rm res} \equiv T_{\rm cal} -T_{\rm model} $)  at 1023 MHz from the combination of all scans. Magenta dots mark the positions of point sources with flux $>1$ Jy at 1.4 GHz, which were not included in the sky model used for calibration.}
\label{fig:sky_map}
\end{figure}

For the calibration of the tracking parts of the TOD, we track the centre and four outskirt pointings of the celestial calibrator source, and compare the raw signals with a model that includes the calibrator flux response at different beam positions. We can obtain the diode power fitting results, $T_{\rm diode}(\nu)$, for two tracking observations that are before and after the scan, respectively. Then, we use the average of the two $T_{\rm diode}(\nu)$ values in the next step to calibrate the scanning data.

 For the calibration of the scanning part, faint point sources in the target field were initially ignored under the assumption that the noise would dominate the residuals (\citealt{2021MNRAS.505.3698W}). Thus \autoref{equ:model} becomes
 \begin{gather}\label{equ:model3}
  T_{\rm model}(t, \nu)= T_{\rm diffuse}(t, \nu) + T_{\rm el}(t, \nu)     +  T_{\rm diode}(t, \nu) 
 + T_{\rm rec} (t, \nu).
\end{gather}
Similar to the calibration for tracking data, we fit the scanning data channel by channel. We obtain the $g(t, \nu)$ and $T_{\rm cal}(t, \nu) $, etc. Then we calculate the sky temperature by
  \begin{gather} \label{eq:Tsky}
T_{\rm sky}(t,\nu)\equiv T_{\rm cal}(t, \nu) - T_{\rm el}(t, \nu) - T_{\rm rec} (t, \nu),
\end{gather}
for diode off time samplings.

We construct equivalent TOD data arrays of total intensity (corresponding to Stokes I), as the mean value of the two calibrated polarisation temperatures, HH and VV. The final maps combine all scans and dishes, after we use the Zenith Equal Area (ZEA) map projection method \citep{2013A&A...558A..33A} to get the coordinates of each TOD point on a pixel grid
with a pixel size of $0.3$ deg chosen to be about 1/3 of the beam size.

With 10.5 hours of data from the MeerKLASS pilot survey undertaken in 2019 (\citealt{2021MNRAS.505.3698W}), the estimated thermal noise of the final data cube, obtained using the ABBA method, was limited to $1.4\,\times$ the theoretical noise level  ($\sim 2$ mK) \textcolor{black}{as estimated using the radiometer equation \citep{2013tra..book.....W}}. 
\textcolor{black}{The ABBA method allows for an estimate of the noise level directly from the calibrated maps. By evaluating the difference between four neighbouring frequency channels, it is possible to isolate the noise contribution, as any signal that is constant or varies linearly across the channels should cancel out.}
\textcolor{black}{In details, the ABBA noise level on one sky pixel is estimated by
\begin{gather}
  \Delta T_i(\nu^{\star})= \frac{1}{2} \left( T^i_{\rm sky}(\nu) + T^i_{\rm sky}(\nu+\delta\nu) \right ) -\frac{1}{2}\left ( T^i_{\rm sky}(\nu-\delta\nu)+T^i_{\rm sky}(\nu+2\delta\nu)\right),
\end{gather}
where the index $i$ goes over all pixels in the map and $^{\star}$ means the combined result from the four channels $[\nu{-}\delta\nu,\,\nu,\,\nu{+}\delta\nu,\,\nu{+}2\delta\nu]$. 
From the observation data, we take a weighted rms,
\begin{gather} \label{equ:rms_obs}
     \Delta T_{\rm RMS}^2(\nu^{\star}) = 
     \frac {N_\text{pix}} {N_\text{pix} - 1}
     \left (\frac {\sum_i w_i\Delta T^2_i} {\sum_i w_i} -
     \left[\frac {\sum_i w_i\Delta T_i} {\sum_i{w_i}}\right]^2 \right),
\end{gather}
where the sum is over the number of pixels in the map, $N_\text{pix}$, and we have suppressed the dependence of $w_i$ and $\Delta T_i$ on $\nu^\star$. For the weight, $w_i$, we used the theoretical expected variance itself, $w_i(\nu^{\star})=1/\sigma^2_{{\rm th},i}(\nu^{\star})$.
The corresponding variance in each pixel would be,
\begin{gather}
\sigma^2_{{\rm th},i}(\nu^{\star})=\frac{1}{4} \Big(\sigma^2_{{\rm th},i}(\nu)+\sigma^2_{{\rm th},i}(\nu{+}\delta\nu)+\sigma^2_{{\rm th},i}(\nu{-}\delta\nu)+ \sigma^2_{{\rm th},i}(\nu{+}2\delta\nu) \Big).
\end{gather}
Here the noise in the initial polarised TOD is given by the radiometer equation 
\begin{equation}
    \sigma^2_{\rm th,pol}(t,\nu) = \frac{T^2_{\rm sys}(t,\nu)}{\delta\nu \, \delta t},
\end{equation}
where $\delta\nu\txteq{=}0.209\,\text{MHz}$ is the frequency width of the MeerKAT L-band channels and $\delta t\txteq{=}2$\,s is the length of each time stamp recording. 
For the system temperature we use the calibrated data itself, $T_{\rm sys}\,{=}\,T_{\rm cal,HH}(t,\nu)$ or $T_{\rm cal,VV}(t,\nu)$  for horizontal and vertical polarisation data, respectively. 
Thus the noise in the initial intensity TOD is
\begin{equation}
    \sigma^2_{\rm th}(t,\nu) = \frac{T^2_{\rm cal,HH}(t,\nu)+T^2_{\rm cal,VV}(t,\nu)}{4\, \delta\nu \, \delta t}.
\end{equation}
In order to get the final data cube, $T^i_{\rm sky}(\nu)$, the data goes through several stages of averaging. We propagate the variance taking into account this averaging in order to get to the final $\sigma^2_{{\rm th},i}(\nu)$.
The theoretical noise level $\sigma_{\rm th} (\nu^{\star}) $ which can be compared to $\Delta T_{\rm RMS} (\nu^{\star})$ in \autoref{equ:rms_obs}, is obtained from the variance
\begin{gather}
     \sigma^2_{\rm th} (\nu^{\star}) = N_\text{pix} \left ( \sum_i \sigma^{-2}_{{\rm th},i}(\nu^{\star})  \right )^{-1}.
\end{gather}
More details can be found in Section 5.6 of \citet{2021MNRAS.505.3698W}. }

We show the final sky temperature map at 1023 MHz, along with the corresponding residual map, in \autoref{fig:sky_map}. The $T_{\rm sky}$ and $T_{\rm res}$ maps from the 64 individual MeerKAT dishes can be found in Figures 24-25 of \cite{2021MNRAS.505.3698W}.

As data quality improved, particularly in the 2021 MeerKLASS deep-field survey (27 repeated scans over $236\,\deg^2$), a more sophisticated iterative self-calibration strategy became necessary. 
This advanced self-calibration method replaces the $T_{\rm diffuse}$ model in \autoref{equ:model3} with the previously calibrated $T_{\rm sky}(t,\nu)$. 
As illustrated in \autoref{fig:pipe}, the procedure is performed iteratively, 
with each loop using the Level-6 $T_{\rm sky}(t,\nu)$ from the previous iteration, 
until the standard deviation of $T_{\rm res}$ converges to a stable value.
By introducing the iterative \textit{self-calibration} process, the estimated thermal noise of the reconstructed maps \textcolor{black}{($\Delta T_{\rm RMS}(\nu^{\star})$)} of the 2021 deep-field survey is limited to ${\sim}\,1.21$\,mK ($1.2\,\times$ the theoretical noise level \textcolor{black}{$\sigma_{\rm th} (\nu^{\star})$}). 
The sky image is shown in Figure 4 of \citet{2025MNRAS.537.3632M}.

\subsection{Synchrotron spectral index}


Diffuse Galactic synchrotron emission is the dominant foreground contaminant for the detection of cosmological ${\rm{H_{I}}}$ at arcmin/degree scale resolutions. Developing a better understanding of this emission is key to both the optimal calibration of experiments which rely on radio sky models, as well designing robust and successful foreground separation/avoidance strategies. In its own right, diffuse Galactic synchrotron emission provides a measurement of our Galactic magnetic field strength and a window on cosmic ray propagation through that magnetic field \citep{orlando13, padovani21, bracco24}, as well as tracing dark matter annihilations \citep{haze, manconi22} and helping to constrain the magnitude of transient and supernova remnant detections \citep{Ocker22, k23}. 

Observing between 544 and 1711 MHz (UHF- through to L-Band) at degree scale resolutions the MeerKAT Large Area Synoptic Survey (MeerKLASS) is optimally placed, in terms of frequency and spatial resolution, to obtain new constraints on the amplitude and spectral dependency of diffuse Galactic synchrotron emission. Additionally, the high spectral resolution of the MeerKLASS data, measurements every 0.2\,MHz for L-Band data, provides a unique continuum science opportunity to follow the frequency dependancy of synchrotron emission without the need to take into account multiple calibration schemes across multiple instruments.

In \cite{mine} the MeerKLASS collaboration use $\sim$ 10 hours of pilot L-Band data between 971 and 1075\,MHz to probe the synchrotron spectral index within the $145^{\circ} < \alpha < 180^{\circ}$, $-1^{\circ} < \delta < 8^{\circ}$ sky region. Synchrotron emission is typically modeled as a power law: 
\begin{equation}
T_\text{sy}(\nu, p) \propto \left(\frac{\nu}{\nu_{0}} \right)^{\beta_\text{sy}(p)},
\end{equation}
with a spectral index which changes across both pixel and frequency. One suggested model for the change in spectral index across frequency, proposed in \cite{arcade}, is as follows:  
\begin{equation}
\beta = {\beta_{0} \, + \, c \, {\rm{ln}}(\nu / \nu_{0})},
\end{equation}
where $\beta_{0}$ is the synchrotron spectral index at frequency $\nu_{0}$ and $c$, often referred to as the curvature term, determines the change to the spectral index over frequency. 

\begin{figure}
\centering
{\includegraphics[width=0.49\linewidth]{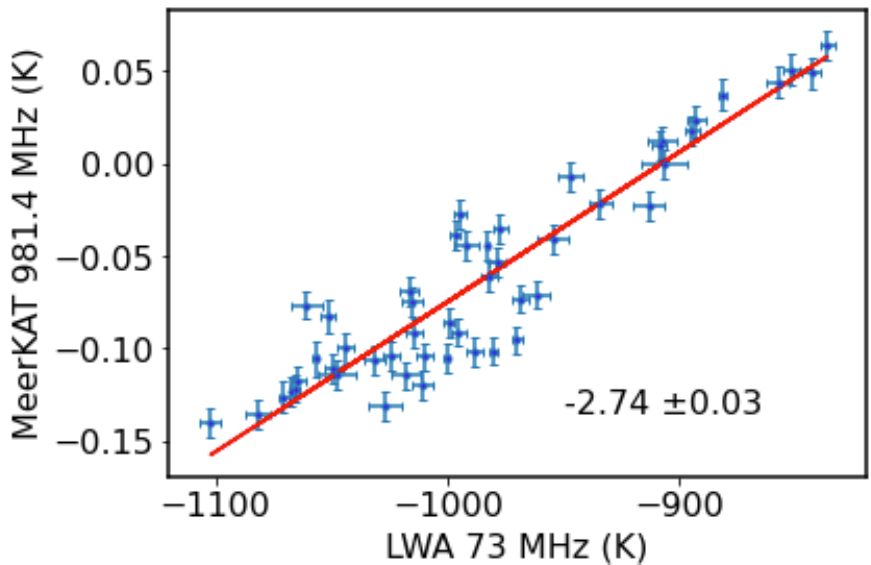}}
{\includegraphics[width=0.49\linewidth]{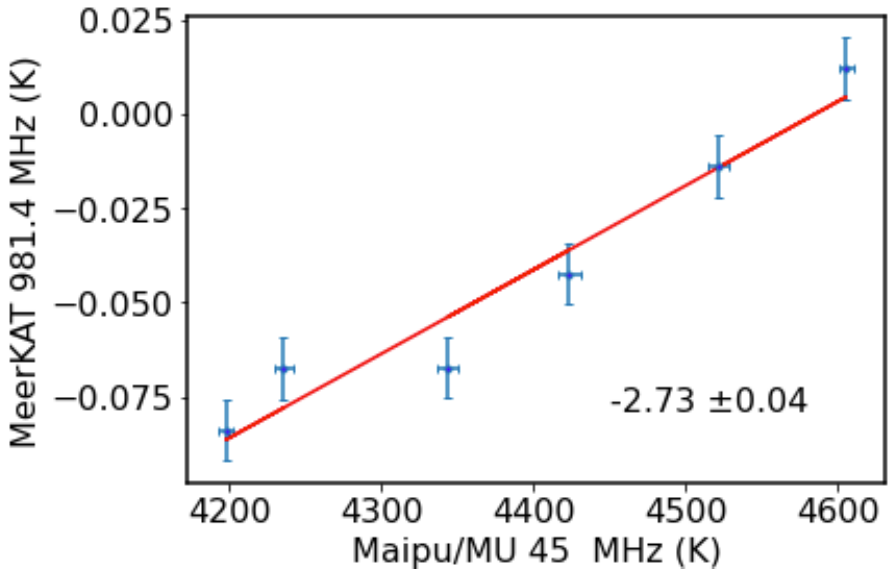}}\\
\caption{Linear regression between MeerKLASS and ancillary data: OVRO-LWA at 1.8$^{\circ}$ resolution ({\emph{left}}) and Maipi/MU at 5$^{\circ}$ resolution ({\emph{right}}).}
 \label{fig:tts}
  \end{figure}

Through the use of linear regression between MeerKLASS and ancillary datasets measurements of the average spectral index across the $154^{\circ} < \alpha < 163^{\circ}$ region were obtained under the assumption that the MeerKLASS maps are so dominated by diffuse Galactic synchrotron emission that, in the absence of any component separation, all other emissions (foreground and cosmological) are negligible. \autoref{fig:tts} shows the average $\beta$ between 73 and 981\,MHz at resolution of $1.8^{\circ}$ alongside the average $\beta$  between 45 and 981\,MHz  at resolution of $5.0^{\circ}$. The OVRO Long Wavelength Array \citep{lwa} provided the 73 MHz data and the Maipu/MU surveys \citep{guz} provided the 45\,MHz data.   
 
In order to calculate the spectral index per pixel between two sky maps both maps require absolute zero-level calibration so that their temperature values only represent the temperature of diffuse Galactic synchrotron and no other contributions such as a receiver temperature or the cosmic microwave background monopole, for instance. \citet{wehus} measure the Haslam 408\, MHz data \citep{oldhas} monopole to be $8.9 \pm 1.3$\,K; subtracting this value from the Haslam all-sky map provides an all-sky template for synchrotron emission which \cite{mine} use, alongside linear regression with their data to produce a cube (R.A, Dec, $\nu$) of MeerKLASS data representing pure synchrotron emission. Fitting a spectral index per pixel across this cube the MeerKLASS collaboration found an average spectral index of $\beta = -2.76 \pm 0.15$ between  971 and 1075\,MHz. 

\begin{figure}
\centering
{\includegraphics[width=0.49\linewidth]{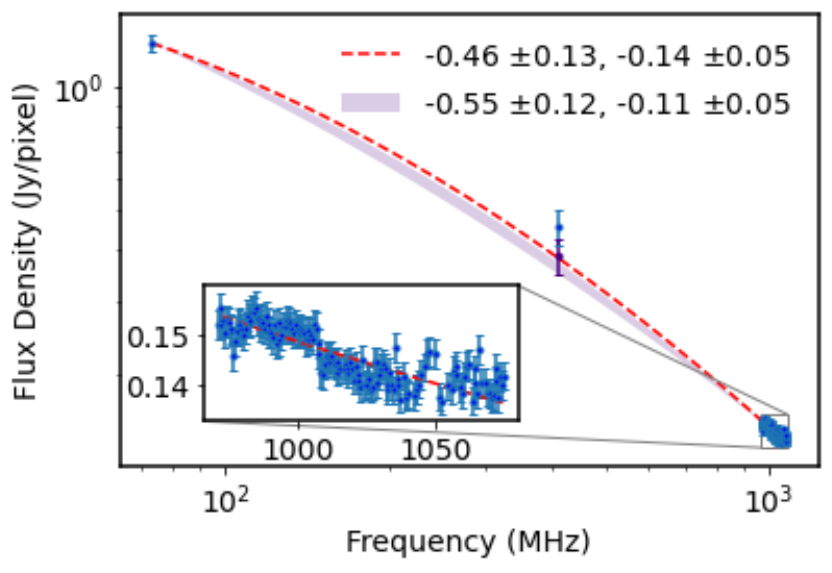}}
{\includegraphics[width=0.452\linewidth]{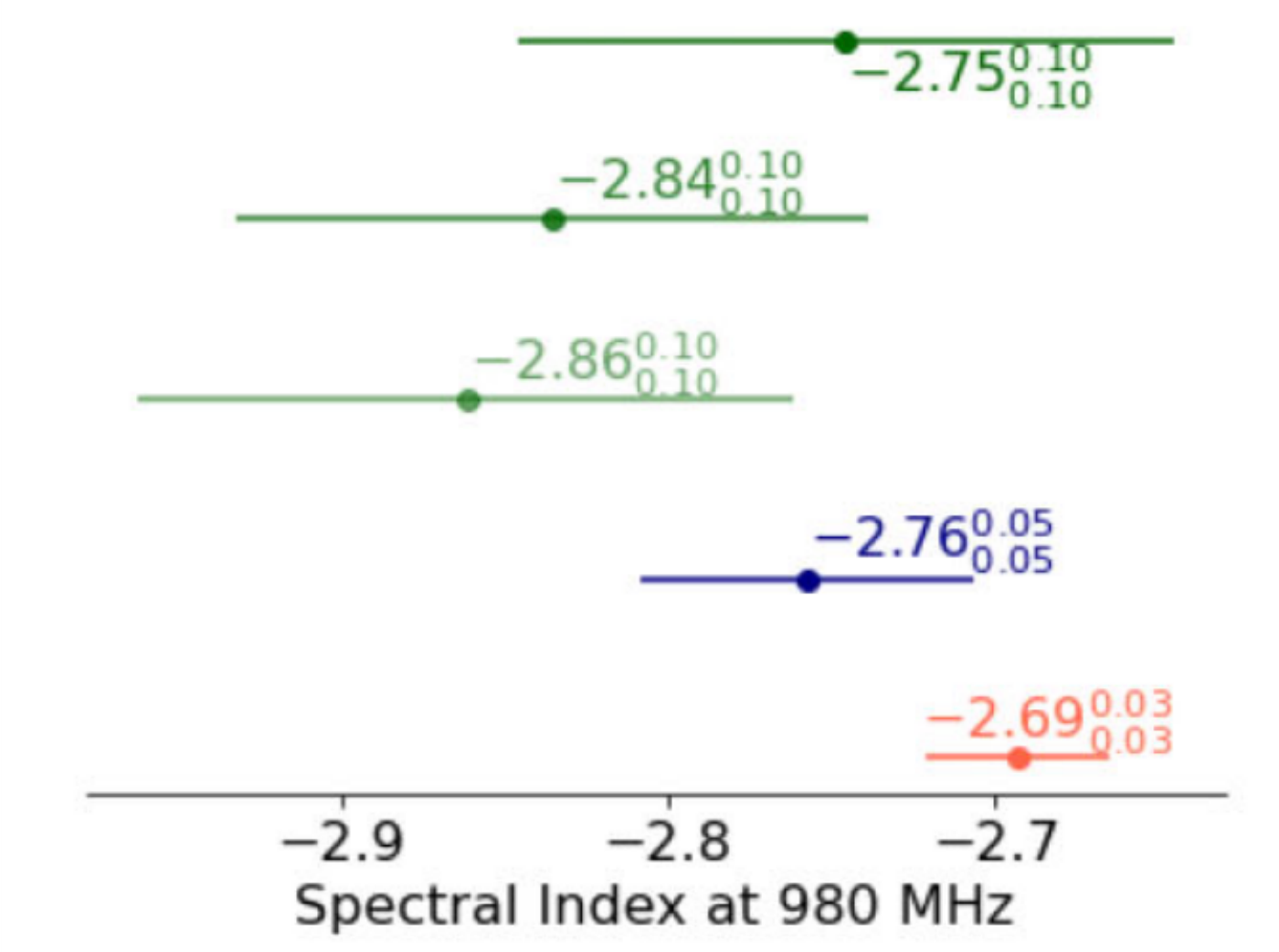}}\\
\caption{Spectral energy distribution between OVRO-LWA, Haslam and MeerKLASS data ({\emph{left}}). MeerKLASS (green), EDGES (blue) and ARCADE2 (red) predictions for the spectral index at 980\,MHz ({\emph{right}}).}
 \label{fig:tts2}
  \end{figure}

Another method to fit spectral forms using multiple maps each with unknown zero-level calibration is aperture photometry which calculates the flux of a point source (within an inner aperture) above the background emission level (outer annulus). \cite{mine} apply a variation on this method, considering instead the average diffuse flux within a 1.8 degree aperture above the minimum map temperature at each frequency. This was done for three aperture regions; an example of the spectral energy distribution for one is shown on the left of \autoref{fig:tts2}, the data point at 73\,MHz is OVRO-LWA data, at 408 MHz the data are provided by Haslam (red dotted fit) and a synchrotron emission model based on Haslam data \citep{planck_CS_2016} (purple fit). The fitted spectral index was found to be $\alpha =  -0.55 \pm 0.12$ at 73\,MHz in units of flux density or $\beta =  -2.55 \pm 0.12$ in units of temperature and the amount of curvature measured was $-0.11 \pm 0.05$. The right-hand plot in \autoref{fig:tts2} shows the comparison between the predicted $\beta$ values at 981\.MHz from the three MeerKLASS apertures alongside predicted $\beta$ values from two other experiments that have placed constraints on the synchrotron spectral index: ARCADE2 \citep{arcade} and EDGES \citep{edges}. For the spectral curvature results, the synchrotron template made from Haslam data was preferred over the Haslam data themselves due to uncertainties over the magnitude of the calibration errors associated with Haslam data. The MeerKLASS data are being used to further probe such gain calibration errors in \cite{mikes}.  
  
\subsection{Foreground cleaning calibrated maps
}\label{sec:cleaning}

At the same frequencies of the redshifted 21\,cm line, other astrophysical sources, such as the diffuse emission from the Milky Way, shine with considerably higher intensities, constituting massive foregrounds to the cosmological signal from \hi. Different strategies have been proposed to tackle the contaminant subtraction problem in this context \citep[see][]{Spinelli01.2026.SKA}. The MeerKLASS collaboration has opted to be as agnostic as possible regarding the nature of the contaminants through the so-called ``Blind Source Separation'' algorithms. In particular, we have employed Principal Component Analysis (PCA). The driving assumption is that astrophysical foregrounds are smooth along frequency, holding few spectral degrees of freedom that can be identified through PCA. The task becomes more complex because extra systematic components, such as bandpass fluctuations, unmodeled beam response, and low-level, undetected RFI, disrupt the foreground smoothness, making the decomposition much more challenging to perform and evaluate.

A noise-weighted PCA was conducted to remove contaminants in the studies described by \citet{Cunnington:2022uzo} and \citet{2025MNRAS.537.3632M}. These analyses achieved a level of cleaning sufficient to detect a cross-correlation signal with overlapping galaxy datasets. However, we encountered two major drawbacks: signal loss, which we addressed using a transfer function approach \citep{TFpaper}, and the ambiguity in selecting the level of cleaning, which increases the variance associated with the final measurement. To address and mitigate these issues, a more optimised cleaning pipeline was necessary.

In \citet{mPCA}, we examined the cleaning problem in greater detail. Specifically, we used the galaxy cross-correlation detection reported by \citet{Cunnington:2022uzo} as a benchmark to test our pre-processing and cleaning choices. We successfully re-detected the cosmological signal by reducing the number of subtracted PCA components from 30 to 4. This achievement reduced signal loss due to cleaning and opened new avenues for modelling and interpreting the identified contaminants.

\citet{mPCA} also developed a new optimised multiscale PCA (mPCA) method. This technique involves splitting intensity maps into large- and small-scale components and performing the cleaning procedure separately for each scale. The cosmological signal and galactic diffuse emissions are closely intertwined at larger scales. However, at smaller scales---although we expect less overlap with the \hi~ signal---the spectral structure of the contaminants can become more complex due to the mixture of systematics and astrophysical foregrounds. We optimise the separation in these different `environments' by allowing the algorithm to operate independently on each scale.

In all tests presented by \citet{mPCA}, mPCA outperformed the standard PCA. We are currently applying mPCA in the latest analysis of MeerKLASS data (see \secref{sec:autocorr}).

\section{Cosmological detections with MeerKLASS single-dish intensity maps}\label{sec:Detections}

A central objective for MeerKLASS has been to demonstrate that 21\,cm intensity mapping can deliver robust cosmological detections using a large multi-dish telescope operated in single-dish mode. Achieving this marks the critical step in establishing the approach as a viable technique for SKA-Mid, and as a cornerstone of its future cosmology programme. Each new detection strengthens confidence in the stability, calibration, and analysis pipeline required for this technique, providing a rigorous end-to-end test of the full observational framework. Through successive campaigns, MeerKLASS has now repeatedly demonstrated this capability, with a series of cosmological detections that refine our understanding of systematic effects and data processing. These results confirm that single-dish 21\,cm intensity mapping is a viable and maturing observational method, laying the foundation for SKAO to employ this mode as a unique resource for cosmological parameter inference across unprecedented cosmic volumes.

\subsection{Cross-correlations with galaxy surveys} \label{sec:xgalaxies}

Overlapping galaxy surveys provide a robust test of data quality for an SKAO intensity mapping precursor like MeerKAT. Due to the novelty of the observations being conducted by MeerKLASS, the data can be prone to systematics, which can be challenging to model or mitigate. Cross-correlating the intensity maps with overlapping galaxy catalogues, which will not share the additive systematics, effectively cuts through to the cosmological components within each tracer and can validate the presence of \hi\ within the cleaned intensity maps.

So far, MeerKLASS has made three successful detections of cosmological \hi\ clustering by detecting a cross-correlation power spectrum with overlapping galaxy surveys. We present these results in \autoref{fig:crossPks}, the full details of which can be found in \citet{Cunnington:2022uzo,mPCA,2025MNRAS.537.3632M}. There is a good consistency between the three power spectra despite them involving different pipelines, survey masks and even data sets for the case of the L-band deep field $\times$ GAMA. We briefly summarise the two distinct surveys and their galaxy catalogue counterparts below:

\begin{figure}
\centering
{\includegraphics[width=0.9\linewidth]{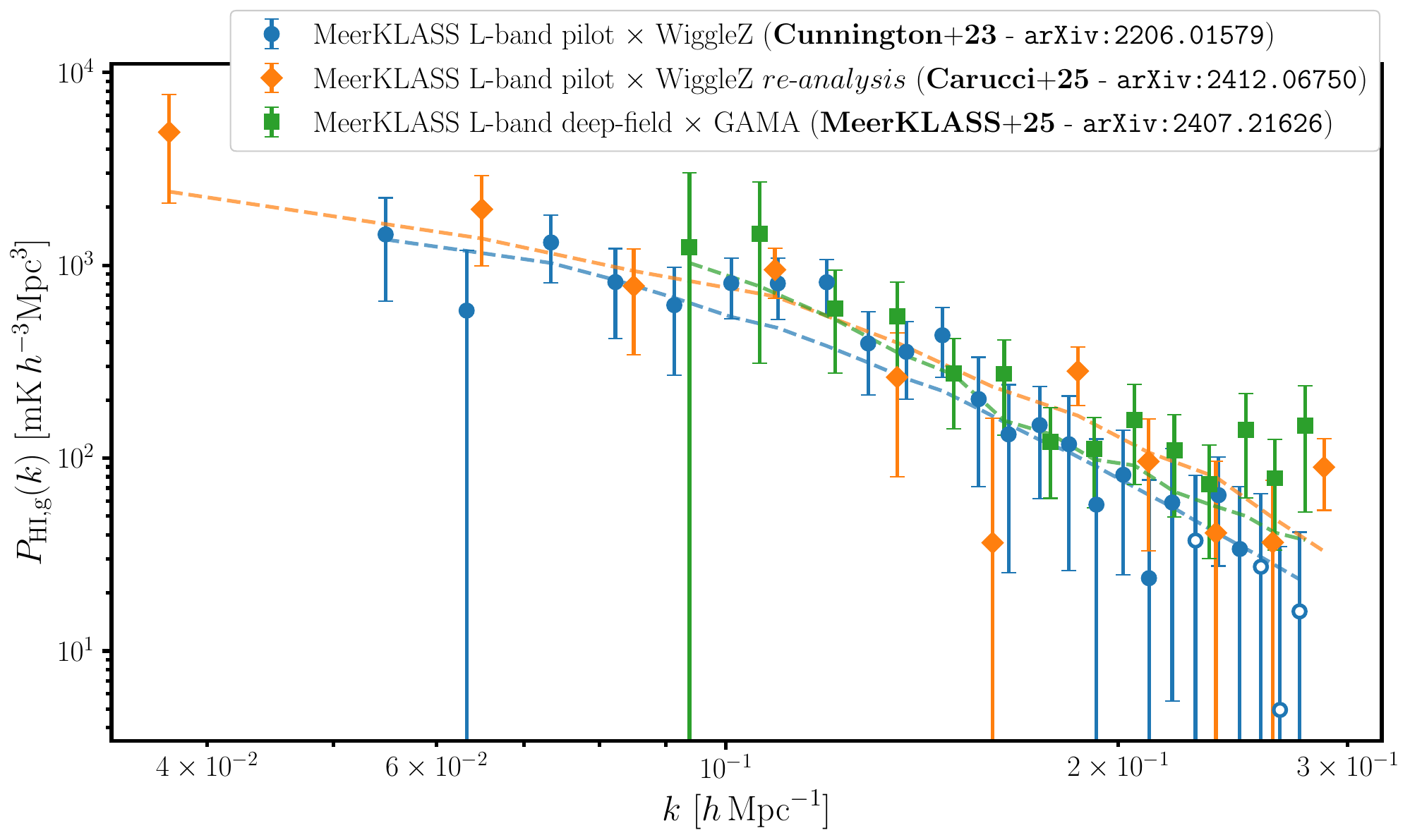}}
\caption{Cross-correlation power spectra between both MeerKLASS L-band intensity maps and overlapping galaxy surveys, all at $z\txteq{\sim}0.43$. Coloured dashed lines are the fitted models to each set of data points. Hollow markers indicate negative power.}
\label{fig:crossPks}
\end{figure}

\textbf{MeerKLASS L-band \textit{pilot} survey $\times$ WiggleZ}: ${\sim}10$ hours of useable observations between $1015{-}973\,$MHz ($0.4\txteq{<}z\txteq{<}0.46$), targeting a single patch of ${\sim}\,200\,\text{deg}^2$ spanning $(153^{\circ},172^{\circ})$ in R.A., and $(-1^{\circ}, -8^{\circ})$ in Declination. This overlapped with the WiggleZ 11hr field, and within this redshift range lay 4031 galaxies \citep{Blake:2010xz,WiggleZ:2018def}. 

\textbf{MeerKLASS L-band \textit{deep-field} survey $\times$ GAMA}:  ${\sim}40$ hours of useable observations, in a very similar frequency interval of the L-band ($1023.6{-}971.2\,$MHz) compared to the pilot survey, providing effectively the same $z\txteq{\sim}0.43$ redshift. This field targeted a single patch of ${\sim}\,200\,\text{deg}^2$ spanning $(330^{\circ},360^{\circ})$ in R.A., and $(-36^{\circ}, -25^{\circ})$ in Declination. This coincided with the Galaxy And Mass Assembly (GAMA) survey and its 23\,hr (G23) field, providing 2269 overlapping galaxies \citep{Driver:2010zb,LiskeGAMA15,Driver:2022vyh}.

Both these data sets have a small ${\sim}\,200\degsq$ area and only span the small RFI-free redshift range of $0.39\txteq{<}z\txteq{<}0.46$. This is therefore only scratching the surface of what will be possible with the full SKA-Mid AA4 Band 1 survey, which should comfortably survey four orders of magnitude larger cosmic volume than achieved with these early MeerKLASS surveys. This opens the possibility for exciting overlap with Stage-IV galaxy surveys, e.g. DESI \citep{DESI:2016fyo}, Euclid \citep{Euclid:2024yrr}, LSST/Rubin \citep{LSSTDarkEnergyScience:2018jkl}, 4MOST \citep{2019Msngr.175....3D}.

\subsection{\hi\ emission from stacking onto galaxy positions}

\begin{figure}
    \centering
    \includegraphics[width=0.33\linewidth]{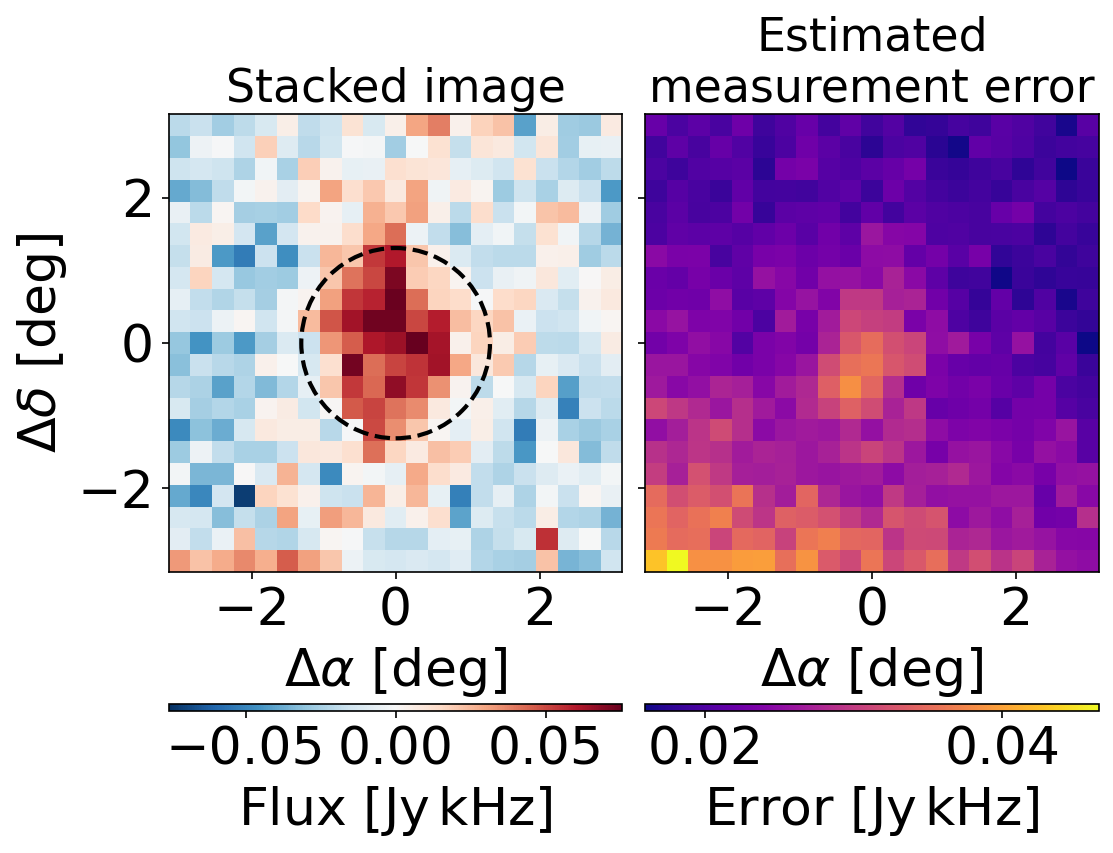}
    \hfill
    \includegraphics[width=0.66\linewidth]{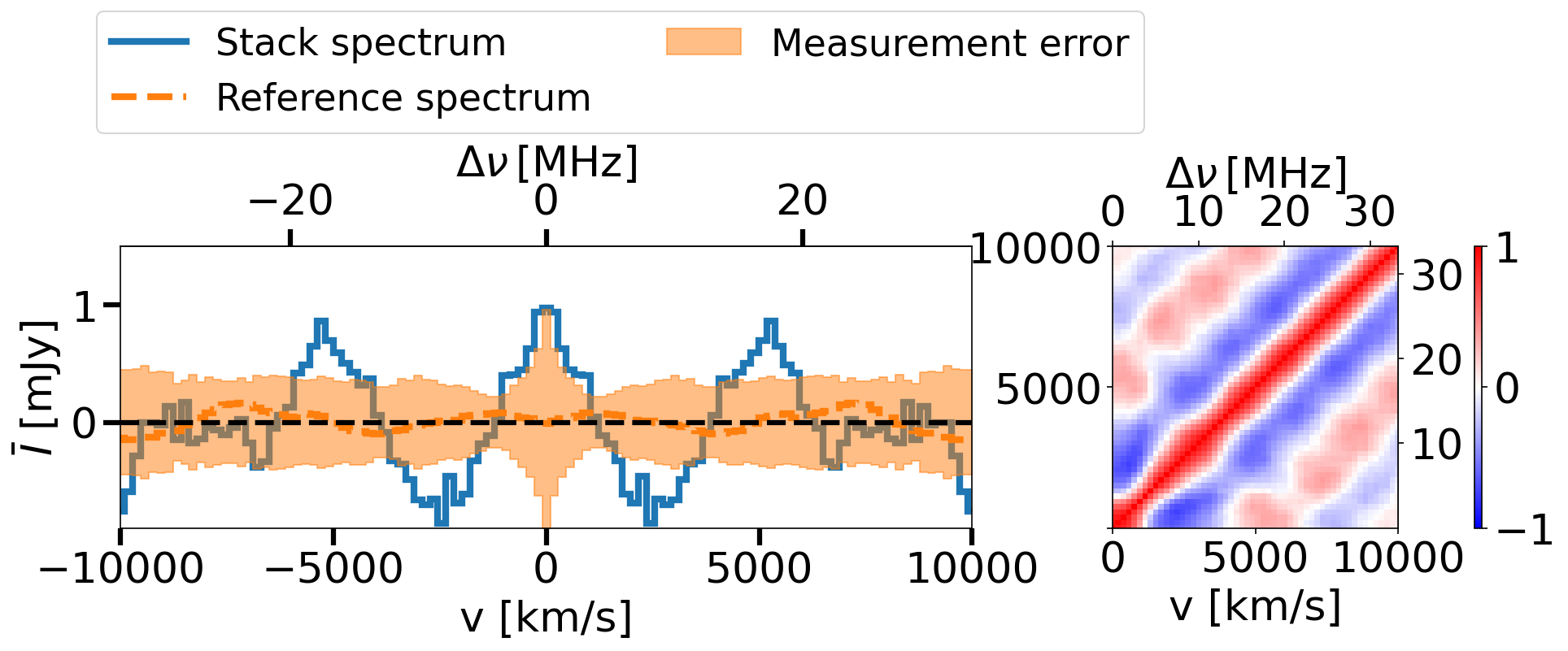}
    \caption{Left panel: The measured angular stacked image. The dashed circle shows the $1.2\,$deg boundary. Central panel: The measurement of the stacked spectrum (``Stacked spectrum''). The orange dashed line shows the average of the reference spectrum over the random shuffles (``Reference spectrum''). The shaded region shows the estimated measurement error (``Measurement error''). Right panel: The estimated correlation matrix for the stacked spectrum.}
    \label{fig:2021lband_stack}
\end{figure}

The stacking of the 21\,cm emission line signal using positions of optical galaxies is a powerful way of probing cosmic \hi, and is one of the main scientific goals of interferometric imaging with SKA-Mid \citep{Sinigaglia:2025weq}. For line intensity mapping surveys, measurements have been made with the \hi\ intensity maps using the Parkes telescope \citep{2019MNRAS.489..385T,2020MNRAS.498.5916T} and the Canadian Hydrogen Intensity Mapping Experiment \citep{2023ApJ...947...16A}, CO emission line signal using the CO Mapping Array Project data \citep{2024ApJ...965....7D}, and Lyman-$\alpha$ intensity maps using the Hobby-Eberly Telescope Dark Energy Experiment observations~\citep{2022ApJ...929...90L, 2022ApJ...934L..26L}.

The single-dish intensity mapping observations, using MeerKAT and future SKA-Mid, pose unique challenges to the stacking measurement due to the low angular resolution of the primary beam. The beam size of MeerKAT, and similarly the SKA-Mid, corresponds to a comoving scale of $\sim 30\,$Mpc in L-band. To understand the detectability of the stacked signal, in \citet{2025arXiv250403908C}, an extensive framework for simulating mock observations was built. It was found that the stacked emission contains a contribution from the source near the centre of the stacking cubelet, as well as a large component of double-counting due to the beam. Foreground removal methods such as PCA can be used to remove the double-counting, while the emission around the centre can be detected at high statistical significance. The stacked signal is much wider along the frequency direction compared to the expected emission line profile of individual \hi galaxies, with a large contribution from the clustering of the \hi sources (see also \citealt{2025arXiv250321743D}). 

Due to the clustering effect, the stacked signal is correlated at different frequency/velocity channels. Due to the smoothing effect from the beam, the stacked signal is correlated in the angular plane. The complexity of the stacked signal requires forward modelling and poses challenges for the covariance estimation. \citet{2025arXiv250403908C} found that the modelling of the covariance is robust if the signal is averaged into a spectrum along the frequency direction and symmetrised along $\Delta\nu = 0$ \citep{2022MNRAS.514.4205S}. The noise covariance can be estimated from the random shuffling of the galaxy positions, whereas the signal covariance can be estimated from forward modelling the stacked signal.

Applying the stacking analysis to the MeerKLASS L-band deep-field data, a stacked \hi\ signal onto the positions of the GAMA galaxies was detected as shown in \autoref{fig:2021lband_stack}. The detection significance is $8.66\sigma$ when averaged into an angular map, and $7.45\sigma$ when averaged into a spectrum. The angular map of the stacked signal shows a clear excess around the centre of the image, corresponding to the size of the primary beam.

The stacked spectrum exhibits an oscillating component of systematics, as seen in \autoref{fig:2021lband_stack}. The frequency of the oscillation matches the characteristic frequency of the primary beam ripple of the MeerKAT telescope \citep{6410347,2021MNRAS.502.2970A}. Using the random shuffling as a null test, we do not find any feature of the systematics, suggesting that the systematics are convolved with the map data. The estimated correlation matrix further demonstrates the convolution.

Using the forward modelling technique, \citet{2025arXiv250403908C} performed Bayesian analysis and constrained the frequency of the oscillation to be $\nu_{\rm sys}={17.90}_{-4.27}^{+6.53}\,$MHz. The parameter fitting gives the effective \hi\ mass of the sources to be $\log_{10}[\langle M_{\hi}\rangle/M_\odot ]=9.84^{+0.48}_{-0.59}$, which is an underestimation. It was found that there is a strong degeneracy between the amplitude of the systematics and the \hi\ density, leading to strong posterior projection effects.

\subsection{The \hi\ auto-correlation power spectrum}
\label{sec:autocorr}


Up to this point, the cosmological measurements presented in this chapter have focused on cross-correlations between the MeerKLASS \hi\ intensity maps and overlapping optical galaxy surveys. However, to fully unlock the potential of \hi\ intensity mapping as a cosmological probe, it is essential to demonstrate that cosmological information can also be recovered from the \hi\ auto-power spectrum. A detection of the \hi\ auto-power has been reported with MeerKAT operating as an interferometer \cite{Paul:2023yrr}, but this measurement probed relatively small, non-linear scales where cosmological parameter inference would be highly complex.

\begin{figure}
    \centering
    \includegraphics[width=0.55\linewidth]{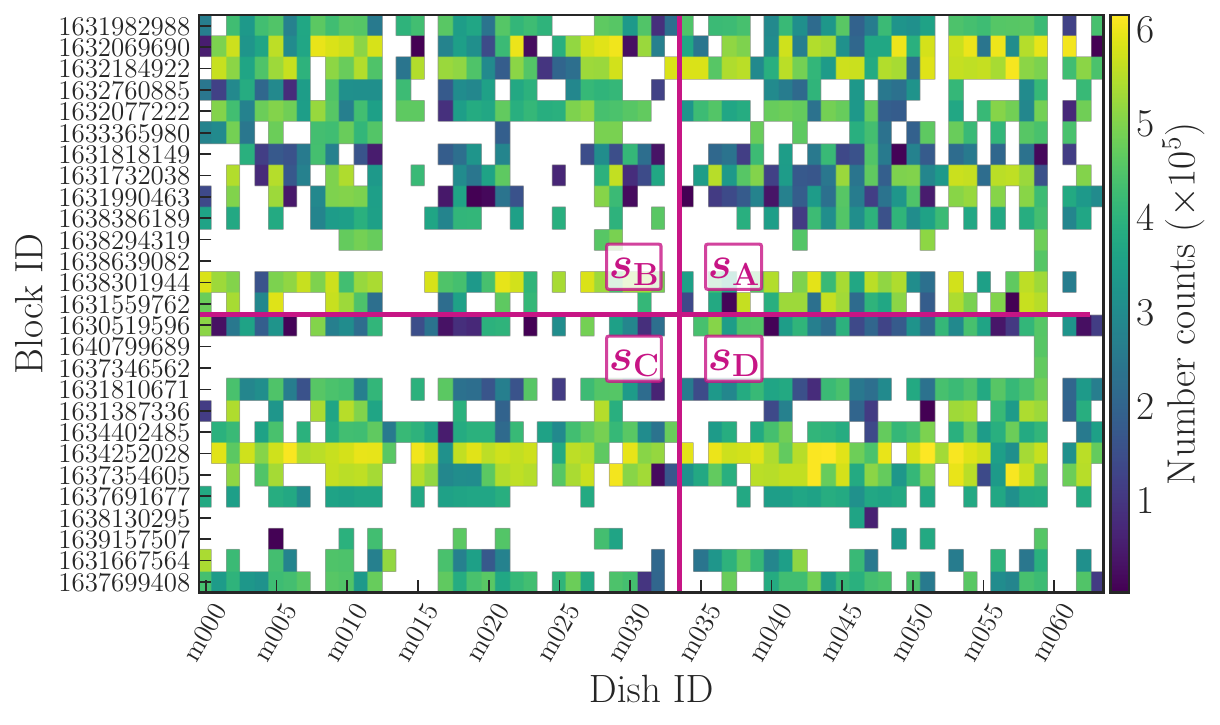}
    \hfill
    \includegraphics[width=0.44\linewidth]{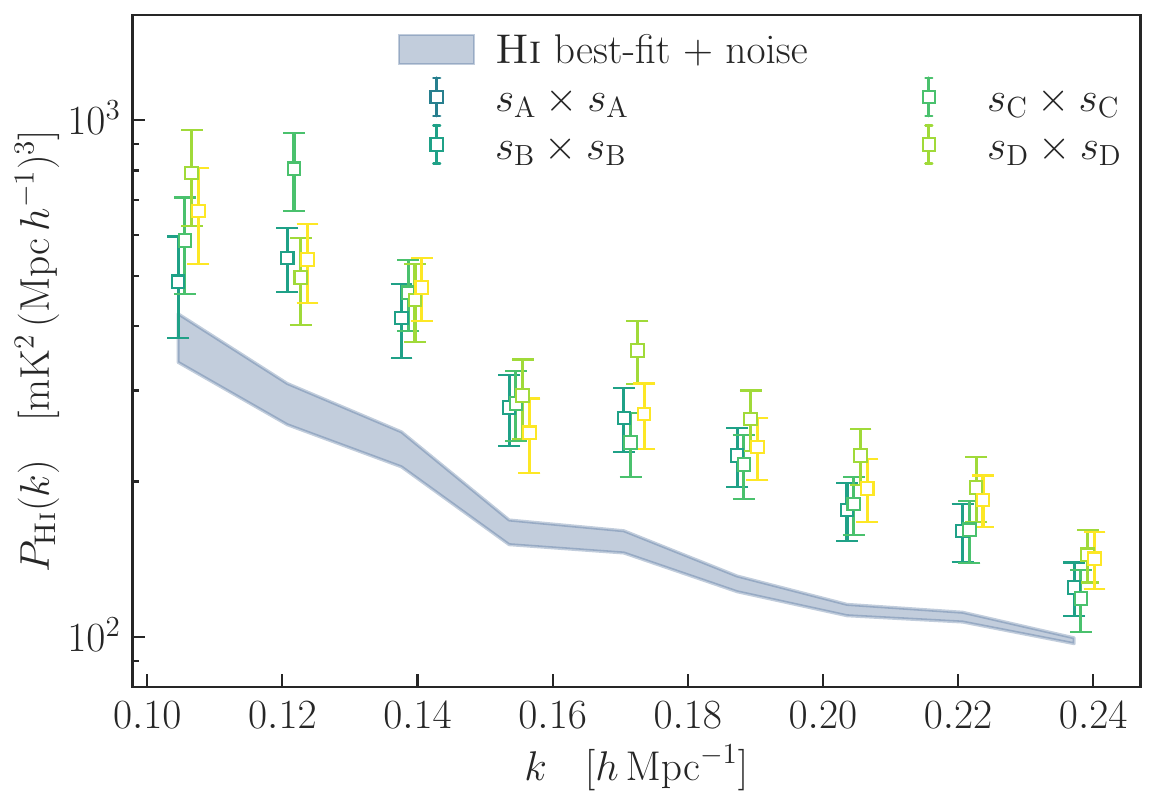}
    \caption{\textit{Left panel:} Number count of useable (unflagged due to contamination) 3D voxels in maps within the 971–1075\,MHz band per dish ($x$-axis) and observation block ($y$-axis). Out of the 41 observed blocks, 14 are completely flagged and therefore not shown in this plot. The definition of the subsets ($s_\mathrm{A}$ to $s_\mathrm{D}$) is highlighted with magenta lines and is carried out, ensuring that the global number counts (and thus the noise level) are comparable in each subset. \textit{Right panel:} Measurements of the four auto-subset $P(k)$ (shown as squares with error bars), compared to the theoretical prediction computed using the best-fit value of $\Ohi\bhi$ obtained from the cross-power spectrum analysis. The estimated noise level \textcolor{black}{(as computed with the ABBA method)} has been added to the theoretical prediction for comparison.}
    \label{fig:2021lband_auto}
\end{figure}

By leveraging the 2021 L-band \textit{deep-field} survey, described in \secref{sec:xgalaxies}, MeerKLASS are honing in on a measurement of the \hi\ power spectrum on cosmological scales, without relying on external data sets. In the absence of a galaxy survey counterpart which avoids additive biases from systematics in the \hi\ data, residual contamination can instead be mitigated by splitting the data set into independent subsets. This way, systematics can be suppressed via internal cross-correlations, as any contamination is expected to be largely uncorrelated across subsets. We defined four subsets through a division that is carried out both scan- and dish-wise (note that the pairs $s_\mathrm{A}-s_\mathrm{C}$ and $s_\mathrm{B}-s_\mathrm{D}$ are the most independent since they do not share dishes or scans), making sure that the signal-to-noise ratio in each of them is equivalent. The data splits into the groups, labelled by $s_i$, are illustrated in the right-hand side of \autoref{fig:2021lband_auto}.

The full analysis of these subsets, which are delivering a strong \hi\ auto-power spectrum measurement, is the focus of the upcoming work in \citet{MatildeAutoPk}, to which we refer the reader. Here we only reveal the auto-correlation between subsets, shown in the right panel of \autoref{fig:2021lband_auto}, which represents an upper limit of the \hi\ cosmological signal. In the auto-$P(k)$ it is not possible to disentangle the cosmological signal from noise and residual systematics\textcolor{black}{. The dominant noise contribution is set by the receiver system temperature, which induces a scale-independent thermal noise term that can be estimated empirically with the ABBA method (see \secref{sec:calibration_Lband}). However, additional contributions may arise from residual contamination. For this reason MeerKLASS} 
is leveraging the cross-subset measurements, i.e. $s_\text{A}\txteq{\times}s_\text{C}$ and $s_\text{B}\txteq{\times}s_\text{D}$ which is proving crucial in mitigating and understanding the remaining additive bias. The ultimate goal is to have a robust independent measurement of the \hi\ cosmological signal at large scales and eventually to constrain the amplitude of the \hi\ power spectrum, which depends on the abundance and clustering properties of the \hi. To this end, a multi-subset formalism \citep{2004MNRAS.347..645P, 2009PhRvL.102b1302S, 2009JCAP...10..007M, 2009MNRAS.397.1348W}, that translates the multi-tracer method to our case, was adopted. In particular, we combined the cross-power spectra in a single data vector, focusing on the $s_\mathrm{A}-s_\mathrm{C}$ and $s_\mathrm{B}-s_\mathrm{D}$ pairs, already identified as the most reliable configurations. We are then running a multi-parametric analysis to constrain the overall amplitude of the power spectra and nuisance parameters to account for potential signal loss due to foreground cleaning.

The foreground cleaning procedure adopted for the auto-correlation measurement in \citet{MatildeAutoPk}, is the multiscale PCA described in \secref{sec:cleaning}, and is applied on each subset independently. The coupling between cosmological signal and foreground modes, which is stronger at the largest radial scales, can induce leakages from the foreground cleaning process, in particular cosmological signal removed alongside foregrounds and therefore lost and residual foregrounds in the cleaned data, which, however, we expect not to be correlated across the different subsets, since the mPCA is performed separately on each of them; this is supported by several robustness tests that were specifically carried out to rule out the presence of spurious contributions to the measurements. Concerning signal loss, we are ensuring that its impact is minor in terms of the current uncertainties; nevertheless, we accounted for it by introducing a phenomenological modelling approach with the addition of dedicated nuisance parameters. Once performed the cleaning, we estimated the spherically averaged auto- and cross-subset power spectra in a scale range tailored to exclude the region mostly plagued by systematics (both residual contaminants and signal loss). 

By demonstrating that \hi\ intensity mapping on cosmological scales is achievable with the MeerKAT telescope, this work is a milestone that paves the way for future surveys with the SKAO. An independent detection of the \hi\ cosmological signal would not only demonstrate the robustness of the technique, but also open a new window on large-scale structure, enabling the breaking of key parameter degeneracies (for instance between the galaxy and \hi\ bias parameters) that currently limit the power of cross-correlation analyses. Such a measurement would therefore allow for a substantial improvement of the constraining power also through full multi-tracer analyses, maximally leveraging the potential of the \hi\ field as an independent and powerful probe of the large-scale structure of the Universe. 

\section{Future observations with MeerKLASS and the SKA-Mid}\label{sec:Future}


After initial observations both in the L-band and the UHF-band, MeerKLASS is now running as a MeerKAT Extra Large Project (XLP). The goal is to reach a total of 2,500 hours on the UHF band covering an area of about 10,000\,deg$^2$, which is most of the Southern Sky away from the Galactic plane. This will allow us to probe cosmology between $0.4\txteq{<}z\txteq{<}1.45$.  Observations are ongoing, and by the end of 2025, MeerKLASS should have close to 800 hours observed in the UHF band, covering about 3,500\,deg$^2$. As discussed in \secref{sec:xgalaxies}, the sky area has strong overlap with DESI, the 4MOST cosmology survey, LSST/Rubin Observatory and Euclid. This will allow for a wealth of cosmological cross-correlation analysis and multi-wavelength studies of galaxy evolution. Main cosmological goals include the measurement of the BAO and RSD both independently and in cross-correlation with spectroscopic galaxy surveys (such as DESI/4MOST). This will, in turn, set constraints on the evolution of dark energy and the growth rate of large-scale structure. The large volumes will also allow us to probe large cosmological scales, past the equality peak, where signatures of primordial non-Gaussianity might be found. The commensal interferometric survey will allow studies of galaxy evolution, cluster science and transient searches. Another 500 hours have been awarded under the XLP for 2026, and the goal is to continue observing all the way to 2028/2029 in order to reach the target of 2,500 hours. 

\begin{figure}
    \centering
    \includegraphics[width=0.85\linewidth]{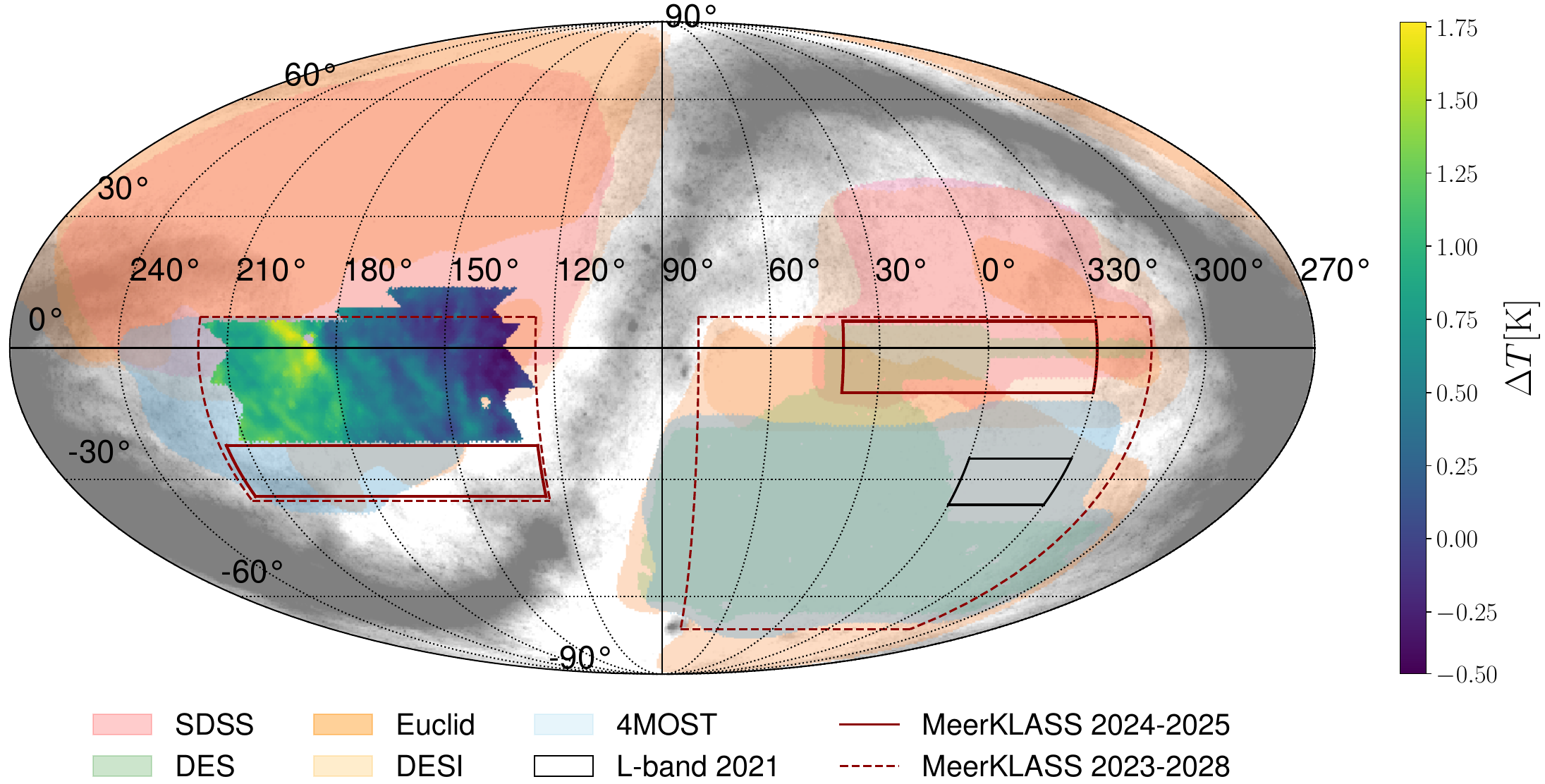}
    \caption{MeerKLASS intensity maps at 760\,MHz shown by the fluctuating colour map region. The full UHF-band range will cover $0.4\,{<}\,z\,{<}\,1.45$ gives it a ${\sim}\,400\times$ larger volume than the L-band deep field (black-solid) ($0.4\,{<}\,z\,{<}\,0.46$) which yielded the most recent published MeerKLASS results. Dotted and solid red outlines mark the approximate footprints for MeerKLASS XLP observations, which commenced in early 2025 and will reach $10{,}000\,\text{deg}^2$. Other shaded regions show some overlapping optical galaxy surveys.}
    \label{fig:Footprint_OpenCall}
\end{figure}

\autoref{table:obs} shows the BAO expected detectability as a function of observing time. \textcolor{black}{Here we assume a 50\% survey efficiency, meaning half of the total observations (including off-scan calibrator tracking) are used in the final data analysis. This is a reasonable assumption from preliminary calibration of MeerKLASS UHF observations. For example, maps centred at R.A., Dec.$\txteq{=}168^\circ ,-2^\circ$ contain 42.4\% of total observations. There is scope for increasing this by implementing a correction to a non-linear gain effect caused by RFI, which is currently responsible for a significant amount of data flagging (this is being pursued in ongoing MeerKLASS work). Hence the 50\% adopted figure remains a reasonable projection from early MeerKLASS findings.} We also refer the reader to the recent review in \citet{Cunnington:2025sdr}, which includes some additional forecasts for the full MeerKLASS survey. 
\autoref{fig:Footprint_OpenCall} shows calibrated UHF band data at 760\,MHz  from the 2022-2023 observation campaign with MeerKLASS and sketched coverage for both the current 2024-2025 and the future ones.

\begin{table}
\centering
\footnotesize
\begin{tabular}{cccccccc}
\hline
{\bf Year}\ & {\bf Obs. time [h]}\  & {\bf Total area [deg$^2$]}\  & {\bf BAO SNR $z$=0.43}\ & {\bf $z$=0.61}\ & {\bf z=0.8}\ & {\bf $z$=1.0}\ & {\bf Combined}\\
\hline
2023 -- 2024* & ~~380 & 1,600 & 1.9& 2.2 & 1.2 & 0.6 & 3.2 \\
 2025~~ & +500 & 3,600 & 2.8 & 3.4 & 1.9 &1.0  & 5.0 \\
2026~~ & +500 & 5,600 & 3.4 &4.4 & 2.4 & 1.3  & 6.2\\
2027~~ & +550 & 7,800 & 4.0 & 5.1 & 2.9 & 1.5  & 7.3\\
2028~~ & +570 & 10,000 & 4.6 & 5.8 & 3.3 & 1.7 & 8.3\\
\hline
\end{tabular}
\caption{Observed and planned XLP survey time and area by year and HI IM BAO wiggles detectability (signal-to-noise ratio - SNR) at each z bin. The first period (*) includes earlier pathfinder observations. The forecast assumes a 50\% survey efficiency. A detection requires SNR $\ge$ 3.} 
\label{table:obs}
\end{table}

\subsection{Extending MeerKLASS strategies to the SKA-Mid}

The transition from MeerKLASS to the SKA-Mid era will represent a natural progression of single-dish intensity-mapping science. Around 2031, the MeerKAT array will be integrated into the SKA-Mid, forging the AA$^*$ operational phase, totalling 144 dishes. The 64 MeerKAT dishes will remain on the same Karoo site as SKA-Mid, ensuring full compatibility in infrastructure, observing conditions, and calibration methods. While the MeerKAT antennas have a 13.5\,m diameter compared with the 15\,m SKA-Mid dishes, this modest difference in collecting area and beam width does not fundamentally alter the observing or calibration approach that has proved successful for MeerKLASS. 
The MeerKAT antennas employ offset-Gregorian optics with cryogenic front ends, closely matching the SKA-Mid design. For SKA-Mid Band 2 (0.95--1.76\,GHz), cryogenic receivers achieve $T_{\rm rec}\txteq{\sim}4$--$7.5\,$K and $T_{\rm sys}\txteq{\sim}14\,$K \citep{8738133}, improving upon MeerKAT’s L-band performance ($T_{\rm sys}\txteq{\sim}16$--$18\,$K). In contrast, Band 1 (0.35--1.05\,GHz) employs ambient-temperature receivers with $T_{\rm rec}\txteq{\sim}15$--$16\,$K, slightly higher than MeerKAT’s cryogenic UHF system ($T_{\rm rec}\txteq{\sim}7\,$K). Both SKA-Mid bands are designed for broad instantaneous bandwidths and high-resolution digital backends (supporting 16k and 65k channel modes) providing enhanced frequency resolution.

Since any differences between MeerKAT and SKA-Mid are marginal, the single-dish scanning and calibration techniques developed by MeerKLASS can be directly adopted by the SKAO. The AA4 deployment, comprising 197 SKA-Mid dishes, will be capable of performing independent auto-correlation (total-power) measurements using a constant-elevation scanning strategy identical to that described in \secref{sec:Calibration}. This approach efficiently mitigates ground-spill and atmospheric fluctuations while building up wide, near-homogeneous sky coverage. The same end-to-end calibration pipeline, anchored on noise-diode injection, regular flux calibration, and iterative self-calibration, will therefore be applicable with minimal modification.

As outlined in \citet{SKA:2018ckk}, SKA-Mid AA4 will aim to conduct a cosmological HI intensity-mapping survey over ${\sim}\,20{,}000\degsq$, covering the redshift range $0.3\txteq{<}z\txteq{<}3$ with Band 1 (350–1050 MHz). This will extend the MeerKLASS concept to cosmological volumes several orders of magnitude larger, exploiting SKA-Mid’s increased expanded frequency coverage. The proven success of MeerKLASS in implementing stable single-dish operation, calibration, and cosmological detections therefore provides the operational and scientific foundation for this next-generation survey, ensuring that SKA-Mid can immediately capitalise on the observing strategies refined with the MeerKAT precursor.

\section{Conclusion}\label{sec:Conclusion}

MeerKLASS has provided the first large-scale demonstration of 21cm intensity mapping with a modern array, establishing a robust framework for cosmological surveys with the SKA-Mid. By combining dedicated calibration strategies, stable scanning operations, and blind foreground-cleaning pipelines, MeerKLASS has achieved the sensitivity and stability required to recover large-scale \hi\ brightness fluctuations from the dominant astrophysical foregrounds. These developments mark a major milestone in proving that single-dish mode observations can deliver the cosmological information required for precision science.

Through the careful validation of calibration pipelines and repeated verification of the single-dish observing mode, MeerKLASS has demonstrated end-to-end control of the systematic effects that once represented the principal barrier to large-scale intensity mapping. The project has achieved multiple cross-correlation detections of cosmological \hi\ fluctuations, placed competitive lower limits on the \hi\ auto-power, and developed rigorous methods to correct for signal loss using the foreground transfer function. These results collectively establish a high degree of readiness for the continuation of large-scale intensity-mapping science with the SKAO.

The lessons learned from MeerKLASS extend directly to the SKAO AA\textsuperscript{*} deployment, where the MeerKAT dishes will be integrated into the SKA-Mid array. Since the same observing mode can be implemented with the expanded system, single-dish intensity mapping represents an ideal early-science verification case for SKA-Mid. The enhanced sensitivity, expanded frequency coverage of Band 1, and extended sky reach to ${\sim}20{,}000\degsq$ will enable SKA-Mid to probe vast cosmological volumes, constraining the evolution of large-scale structure and testing fundamental physics on horizon scales.

Looking ahead, the experience gained from MeerKLASS forms a cornerstone for the SKAO’s large-scale cosmology programme. The techniques and analyses developed here, e.g. calibration stability, data combination, foreground mitigation, and power-spectrum analysis, will ensure that the SKAO can exploit its full potential as a precision cosmological instrument. In this sense, MeerKLASS is not only a scientific precursor but a technological and methodological blueprint, demonstrating that 21cm single-dish intensity mapping has matured into a reliable probe ready for the SKAO era.

\section*{Author list ordering}

The construction of this chapter was jointly led by Steven Cunnington and Jingying Wang, who appear as alphabetical joint first authors. Mario G. Santos is PI of the MeerKLASS experiment. All other authors are listed alphabetically and were chosen as those who made contributions (either directly or in a supervisory capacity) to the results presented in this work. 

\bibliographystyle{abbrvnat-maxbibnames4}
\bibliography{chapter} 

@ARTICLE{2021MNRAS.505.3698W,
       author = {{Wang}, Jingying and {Santos}, Mario G. and {Bull}, Philip and {Grainge}, Keith and {Cunnington}, Steven and {Fonseca}, Jos{\'e} and {Irfan}, Melis O. and {Li}, Yichao and {Pourtsidou}, Alkistis and {Soares}, Paula S. and {Spinelli}, Marta and {Bernardi}, Gianni and {Engelbrecht}, Brandon},
        title = "{H I intensity mapping with MeerKAT: calibration pipeline for multidish autocorrelation observations}",
      journal = {MNRAS},
         year = 2021,
        month = aug,
       volume = {505},
       number = {3},
        pages = {3698-3721},
          doi = {10.1093/mnras/stab1365},
archivePrefix = {arXiv},
       eprint = {2011.13789},
 primaryClass = {astro-ph.CO},
       adsurl = {https://ui.adsabs.harvard.edu/abs/2021MNRAS.505.3698W}
}

@article{Cunnington:2022uzo,
    author = "Cunnington, Steven and others",
    title = "{H\,i intensity mapping with MeerKAT: power spectrum detection in cross-correlation with WiggleZ galaxies}",
    eprint = "2206.01579",
    archivePrefix = "arXiv",
    primaryClass = "astro-ph.CO",
    doi = "10.1093/mnras/stac3060",
    journal = "Mon. Not. Roy. Astron. Soc.",
    volume = "518",
    number = "4",
    pages = "6262--6272",
    year = "2022"
}

@ARTICLE{2017A&C....18....8A,
       author = {{Akeret}, J. and {Seehars}, S. and {Chang}, C. and {Monstein}, C. and
         {Amara}, A. and {Refregier}, A.},
        title = "{HIDE \&amp; SEEK: End-to-end packages to simulate and process radio survey data}",
      journal = {Astronomy and Computing},
         year = 2017,
        month = jan,
       volume = {18},
        pages = {8-17},
          doi = {10.1016/j.ascom.2016.11.001},
archivePrefix = {arXiv},
       eprint = {1607.07443},
 primaryClass = {astro-ph.IM},
       adsurl = {https://ui.adsabs.harvard.edu/abs/2017A&C....18....8A}
}

@ARTICLE{2010MNRAS.405..155O,
       author = {{Offringa}, A.~R. and {de Bruyn}, A.~G. and {Biehl}, M. and
         {Zaroubi}, S. and {Bernardi}, G. and {Pandey}, V.~N.},
        title = "{Post-correlation radio frequency interference classification methods}",
      journal = {MNRAS},
         year = 2010,
        month = jun,
       volume = {405},
       number = {1},
        pages = {155-167},
          doi = {10.1111/j.1365-2966.2010.16471.x},
archivePrefix = {arXiv},
       eprint = {1002.1957},
 primaryClass = {astro-ph.IM},
       adsurl = {https://ui.adsabs.harvard.edu/abs/2010MNRAS.405..155O}
}

@ARTICLE{2013A&A...558A..33A,
       author = {{Astropy Collaboration} and {Robitaille}, Thomas P. and
         {Tollerud}, Erik J. and {Greenfield}, Perry and {Droettboom}, Michael and
         {Bray}, Erik and {Aldcroft}, Tom and {Davis}, Matt and
         {Ginsburg}, Adam and {Price-Whelan}, Adrian M. and
         {Kerzendorf}, Wolfgang E. and {Conley}, Alexander and {Crighton}, Neil and
         {Barbary}, Kyle and {Muna}, Demitri and {Ferguson}, Henry and
         {Grollier}, Fr{\'e}d{\'e}ric and {Parikh}, Madhura M. and
         {Nair}, Prasanth H. and {Unther}, Hans M. and {Deil}, Christoph and
         {Woillez}, Julien and {Conseil}, Simon and {Kramer}, Roban and
         {Turner}, James E.~H. and {Singer}, Leo and {Fox}, Ryan and
         {Weaver}, Benjamin A. and {Zabalza}, Victor and {Edwards}, Zachary I. and
         {Azalee Bostroem}, K. and {Burke}, D.~J. and {Casey}, Andrew R. and
         {Crawford}, Steven M. and {Dencheva}, Nadia and {Ely}, Justin and
         {Jenness}, Tim and {Labrie}, Kathleen and {Lim}, Pey Lian and
         {Pierfederici}, Francesco and {Pontzen}, Andrew and {Ptak}, Andy and
         {Refsdal}, Brian and {Servillat}, Mathieu and {Streicher}, Ole},
        title = "{Astropy: A community Python package for astronomy}",
      journal = {Astronomy \& Astrophysics},
         year = 2013,
        month = oct,
       volume = {558},
          eid = {A33},
        pages = {A33},
          doi = {10.1051/0004-6361/201322068},
archivePrefix = {arXiv},
       eprint = {1307.6212},
 primaryClass = {astro-ph.IM},
       adsurl = {https://ui.adsabs.harvard.edu/abs/2013A&A...558A..33A}
}

@ARTICLE{LiskeGAMA15,
       author = {{Liske}, J. and {Baldry}, I.~K. and {Driver}, S.~P. and {Tuffs}, R.~J. and {Alpaslan}, M. and {Andrae}, E. and {Brough}, S. and {Cluver}, M.~E. and {Grootes}, M.~W. and {Gunawardhana}, M.~L.~P. and {Kelvin}, L.~S. and {Loveday}, J. and {Robotham}, A.~S.~G. and {Taylor}, E.~N. and {Bamford}, S.~P. and {Bland-Hawthorn}, J. and {Brown}, M.~J.~I. and {Drinkwater}, M.~J. and {Hopkins}, A.~M. and {Meyer}, M.~J. and {Norberg}, P. and {Peacock}, J.~A. and {Agius}, N.~K. and {Andrews}, S.~K. and {Bauer}, A.~E. and {Ching}, J.~H.~Y. and {Colless}, M. and {Conselice}, C.~J. and {Croom}, S.~M. and {Davies}, L.~J.~M. and {De Propris}, R. and {Dunne}, L. and {Eardley}, E.~M. and {Ellis}, S. and {Foster}, C. and {Frenk}, C.~S. and {H{\"a}u{\ss}ler}, B. and {Holwerda}, B.~W. and {Howlett}, C. and {Ibarra}, H. and {Jarvis}, M.~J. and {Jones}, D.~H. and {Kafle}, P.~R. and {Lacey}, C.~G. and {Lange}, R. and {Lara-L{\'o}pez}, M.~A. and {L{\'o}pez-S{\'a}nchez}, {\'A}. R. and {Maddox}, S. and {Madore}, B.~F. and {McNaught-Roberts}, T. and {Moffett}, A.~J. and {Nichol}, R.~C. and {Owers}, M.~S. and {Palamara}, D. and {Penny}, S.~J. and {Phillipps}, S. and {Pimbblet}, K.~A. and {Popescu}, C.~C. and {Prescott}, M. and {Proctor}, R. and {Sadler}, E.~M. and {Sansom}, A.~E. and {Seibert}, M. and {Sharp}, R. and {Sutherland}, W. and {V{\'a}zquez-Mata}, J.~A. and {van Kampen}, E. and {Wilkins}, S.~M. and {Williams}, R. and {Wright}, A.~H.},
        title = "{Galaxy And Mass Assembly (GAMA): end of survey report and data release 2}",
      journal = {MNRAS},
         year = 2015,
        month = sep,
       volume = {452},
       number = {2},
        pages = {2087-2126},
          doi = {10.1093/mnras/stv1436},
archivePrefix = {arXiv},
       eprint = {1506.08222},
 primaryClass = {astro-ph.GA},
       adsurl = {https://ui.adsabs.harvard.edu/abs/2015MNRAS.452.2087L}
}

@article{Driver:2022vyh,
    author = "Driver, Simon P. and others",
    title = "{Galaxy And Mass Assembly (GAMA): Data Release 4 and the z \ensuremath{<} 0.1 total and z \ensuremath{<} 0.08 morphological galaxy stellar mass functions}",
    eprint = "2203.08539",
    archivePrefix = "arXiv",
    primaryClass = "astro-ph.GA",
    doi = "10.1093/mnras/stac472",
    journal = "Mon. Not. Roy. Astron. Soc.",
    volume = "513",
    number = "1",
    pages = "439--467",
    year = "2022"
}

@article{WiggleZ:2018def,
    author = "Drinkwater, Michael J. and others",
    collaboration = "WiggleZ",
    title = "{The WiggleZ Dark Energy Survey: final data release and the metallicity of UV-luminous galaxies}",
    eprint = "1910.08284",
    archivePrefix = "arXiv",
    primaryClass = "astro-ph.GA",
    doi = "10.1093/mnras/stx2963",
    journal = "Mon. Not. Roy. Astron. Soc.",
    volume = "474",
    number = "3",
    pages = "4151--4168",
    year = "2018"
}

@article{Blake:2010xz,
    author = "Blake, Chris and others",
    title = "{The WiggleZ Dark Energy Survey: the selection function and z=0.6 galaxy power spectrum}",
    eprint = "1003.5721",
    archivePrefix = "arXiv",
    primaryClass = "astro-ph.CO",
    doi = "10.1111/j.1365-2966.2010.16747.x",
    journal = "Mon. Not. Roy. Astron. Soc.",
    volume = "406",
    pages = "803--821",
    year = "2010"
}

@article{Driver:2010zb,
    author = "Driver, S. P. and others",
    title = "{Galaxy and Mass Assembly (GAMA): survey diagnostics and core data release}",
    eprint = "1009.0614",
    archivePrefix = "arXiv",
    primaryClass = "astro-ph.CO",
    doi = "10.1111/j.1365-2966.2010.18188.x",
    journal = "Mon. Not. Roy. Astron. Soc.",
    volume = "413",
    pages = "971",
    year = "2011"
}

@ARTICLE{2025MNRAS.537.3632M,
       author = {{MeerKLASS Collaboration} and {Barberi-Squarotti}, M. and {Bernal}, J.~L. and {Bull}, P. and {Camera}, S. and {Carucci}, I.~P. and {Chen}, Z. and {Cunnington}, S. and {Engelbrecht}, B.~N. and {Fonseca}, J. and {Grainge}, K. and {Irfan}, M.~O. and {Li}, Y. and {Mazumder}, A. and {Paul}, S. and {Pourtsidou}, A. and {Santos}, M.~G. and {Spinelli}, M. and {Wang}, J. and {Witzemann}, A. and {Wolz}, L.},
        title = "{MeerKLASS L-band deep-field intensity maps: entering the H I dominated regime}",
      journal = {MNRAS},
         year = 2025,
        month = mar,
       volume = {537},
       number = {4},
        pages = {3632-3661},
          doi = {10.1093/mnras/staf195},
archivePrefix = {arXiv},
       eprint = {2407.21626},
 primaryClass = {astro-ph.CO},
       adsurl = {https://ui.adsabs.harvard.edu/abs/2025MNRAS.537.3632M}
}

@ARTICLE{orlando13,
       author = {{Orlando}, Elena and {Strong}, Andrew},
        title = "{Galactic synchrotron emission with cosmic ray propagation models}",
      journal = {MNRAS},
         year = 2013,
        month = dec,
       volume = {436},
       number = {3},
        pages = {2127-2142},
          doi = {10.1093/mnras/stt1718},
archivePrefix = {arXiv},
       eprint = {1309.2947},
 primaryClass = {astro-ph.GA},
       adsurl = {https://ui.adsabs.harvard.edu/abs/2013MNRAS.436.2127O}
}

@ARTICLE{padovani21,
       author = {{Padovani}, Marco and {Bracco}, Andrea and {Jeli{\'c}}, Vibor and {Galli}, Daniele and {Bellomi}, Elena},
        title = "{Spectral index of synchrotron emission: insights from the diffuse and magnetised interstellar medium}",
      journal = {arXiv e-prints},
         year = 2021,
        month = jun,
          eid = {arXiv:2106.10929},
        pages = {arXiv:2106.10929},
archivePrefix = {arXiv},
       eprint = {2106.10929},
 primaryClass = {astro-ph.HE},
       adsurl = {https://ui.adsabs.harvard.edu/abs/2021arXiv210610929P}
}

@ARTICLE{bracco24,
       author = {{Bracco}, Andrea and {Padovani}, Marco and {Galli}, Daniele},
        title = "{A new analytical model of the cosmic-ray energy flux for Galactic diffuse radio emission}",
      journal = {Astronomy \& Astrophysics},
         year = 2024,
        month = jun,
       volume = {686},
          eid = {A52},
        pages = {A52},
          doi = {10.1051/0004-6361/202449625},
archivePrefix = {arXiv},
       eprint = {2402.19367},
 primaryClass = {astro-ph.GA},
       adsurl = {https://ui.adsabs.harvard.edu/abs/2024A&A...686A..52B}
}

@ARTICLE{haze,
       author = {{Hooper}, Dan and {Finkbeiner}, Douglas P. and {Dobler}, Gregory},
        title = "{Possible evidence for dark matter annihilations from the excess microwave emission around the center of the Galaxy seen by the Wilkinson Microwave Anisotropy Probe}",
      journal = {Physical Review D},
         year = 2007,
        month = oct,
       volume = {76},
       number = {8},
          eid = {083012},
        pages = {083012},
          doi = {10.1103/PhysRevD.76.083012},
archivePrefix = {arXiv},
       eprint = {0705.3655},
 primaryClass = {astro-ph},
       adsurl = {https://ui.adsabs.harvard.edu/abs/2007PhRvD..76h3012H}
}

@ARTICLE{manconi22,
       author = {{Manconi}, Silvia and {Cuoco}, Alessandro and {Lesgourgues}, Julien},
        title = "{Dark Matter Constraints from Planck Observations of the Galactic Polarized Synchrotron Emission}",
      journal = {Physical Review Letters},
         year = 2022,
        month = sep,
       volume = {129},
       number = {11},
          eid = {111103},
        pages = {111103},
          doi = {10.1103/PhysRevLett.129.111103},
archivePrefix = {arXiv},
       eprint = {2204.04232},
 primaryClass = {astro-ph.HE},
       adsurl = {https://ui.adsabs.harvard.edu/abs/2022PhRvL.129k1103M}
}

@ARTICLE{Ocker22,
       author = {{Ocker}, Stella Koch and {Cordes}, James M. and {Chatterjee}, Shami and {Gorsuch}, Miranda R.},
        title = "{Radio Scattering Horizons for Galactic and Extragalactic Transients}",
      journal = {APJ},
         year = 2022,
        month = jul,
       volume = {934},
       number = {1},
          eid = {71},
        pages = {71},
          doi = {10.3847/1538-4357/ac75ba},
archivePrefix = {arXiv},
       eprint = {2203.16716},
 primaryClass = {astro-ph.GA},
       adsurl = {https://ui.adsabs.harvard.edu/abs/2022ApJ...934...71O}
}

@ARTICLE{k23,
       author = {{Khabibullin}, I.~I. and {Churazov}, E.~M. and {Bykov}, A.~M. and {Chugai}, N.~N. and {Sunyaev}, R.~A.},
        title = "{SRG/eROSITA discovery of a radio-faint X-ray candidate supernova remnant SRGe J003602.3+605421 = G121.1-1.9}",
      journal = {MNRAS},
         year = 2023,
        month = jun,
       volume = {521},
       number = {4},
        pages = {5536-5556},
          doi = {10.1093/mnras/stad818},
archivePrefix = {arXiv},
       eprint = {2207.00064},
 primaryClass = {astro-ph.HE},
       adsurl = {https://ui.adsabs.harvard.edu/abs/2023MNRAS.521.5536K}
}

@ARTICLE{mine,
       author = {{Irfan}, Melis O. and {Bull}, Philip and {Santos}, Mario G. and {Wang}, Jingying and {Grainge}, Keith and {Li}, Yichao and {Carucci}, Isabella P. and {Spinelli}, Marta and {Cunnington}, Steven},
        title = "{Measurements of the diffuse Galactic synchrotron spectral index and curvature from MeerKLASS pilot data}",
      journal = {MNRAS},
         year = 2022,
        month = feb,
       volume = {509},
       number = {4},
        pages = {4923-4939},
          doi = {10.1093/mnras/stab3346},
archivePrefix = {arXiv},
       eprint = {2111.08517},
 primaryClass = {astro-ph.GA},
       adsurl = {https://ui.adsabs.harvard.edu/abs/2022MNRAS.509.4923I}
}

@ARTICLE{arcade,
       author = {{Fixsen}, D.~J. and {Kogut}, A. and {Levin}, S. and {Limon}, M. and
         {Lubin}, P. and {Mirel}, P. and {Seiffert}, M. and {Singal}, J. and
         {Wollack}, E. and {Villela}, T. and {Wuensche}, C.~A.},
        title = "{ARCADE 2 Measurement of the Absolute Sky Brightness at 3-90 GHz}",
      journal = {APJ},
         year = 2011,
        month = jun,
       volume = {734},
       number = {1},
          eid = {5},
        pages = {5},
          doi = {10.1088/0004-637X/734/1/5},
archivePrefix = {arXiv},
       eprint = {0901.0555},
 primaryClass = {astro-ph.CO},
       adsurl = {https://ui.adsabs.harvard.edu/abs/2011ApJ...734....5F}
}

@ARTICLE{lwa,
       author = {{Eastwood}, Michael W. and {Anderson}, Marin M. and {Monroe}, Ryan M. and {Hallinan}, Gregg and {Barsdell}, Benjamin R. and {Bourke}, Stephen A. and {Clark}, M.~A. and {Ellingson}, Steven W. and {Dowell}, Jayce and {Garsden}, Hugh and {Greenhill}, Lincoln J. and {Hartman}, Jacob M. and {Kocz}, Jonathon and {Lazio}, T. Joseph W. and {Price}, Danny C. and {Schinzel}, Frank K. and {Taylor}, Gregory B. and {Vedantham}, Harish K. and {Wang}, Yuankun and {Woody}, David P.},
        title = "{The Radio Sky at Meter Wavelengths: m-mode Analysis Imaging with the OVRO-LWA}",
      journal = {The Astronomical Journal},
         year = 2018,
        month = jul,
       volume = {156},
       number = {1},
          eid = {32},
        pages = {32},
          doi = {10.3847/1538-3881/aac721},
archivePrefix = {arXiv},
       eprint = {1711.00466},
 primaryClass = {astro-ph.IM},
       adsurl = {https://ui.adsabs.harvard.edu/abs/2018AJ....156...32E}
}

@ARTICLE{guz,
       author = {{Guzm{\'a}n}, A.~E. and {May}, J. and {Alvarez}, H. and {Maeda}, K.},
        title = "{All-sky Galactic radiation at 45 MHz and spectral index between 45 and 408 MHz}",
      journal = {Astronomy \& Astrophysics},
         year = 2011,
        month = jan,
       volume = {525},
          eid = {A138},
        pages = {A138},
          doi = {10.1051/0004-6361/200913628},
archivePrefix = {arXiv},
       eprint = {1011.4298},
 primaryClass = {astro-ph.GA},
       adsurl = {https://ui.adsabs.harvard.edu/abs/2011A&A...525A.138G}
}

@ARTICLE{oldhas,
       author = {{Haslam}, C.~G.~T. and {Salter}, C.~J. and {Stoffel}, H. and {Wilson}, W.~E.},
        title = "{A 408 MHz all-sky continuum survey. II. The atlas of contour maps.}",
      journal = {Astronomy \& Astrophysicss},
         year = 1982,
        month = jan,
       volume = {47},
        pages = {1-143},
       adsurl = {https://ui.adsabs.harvard.edu/abs/1982A&AS...47....1H}
}

@ARTICLE{edges,
       author = {{Mozdzen}, T.~J. and {Mahesh}, N. and {Monsalve}, R.~A. and {Rogers}, A.~E.~E. and {Bowman}, J.~D.},
        title = "{Spectral index of the diffuse radio background between 50 and 100 MHz}",
      journal = {MNRAS},
         year = 2019,
        month = mar,
       volume = {483},
       number = {4},
        pages = {4411-4423},
          doi = {10.1093/mnras/sty3410},
archivePrefix = {arXiv},
       eprint = {1812.02660},
 primaryClass = {astro-ph.IM},
       adsurl = {https://ui.adsabs.harvard.edu/abs/2019MNRAS.483.4411M}
}

@ARTICLE{wehus,
       author = {{Wehus}, I.~K. and {Fuskeland}, U. and {Eriksen}, H.~K. and {Band
        ay}, A.~J. and {Dickinson}, C. and {Ghosh}, T. and {G{\'o}rski}, K.~M. and
         {Lawrence}, C.~R. and {Leahy}, J.~P. and {Maino}, D. and {Reich}, P. and
         {Reich}, W.},
        title = "{Monopole and dipole estimation for multi-frequency sky maps by linear regression}",
      journal = {Astronomy \& Astrophysics},
         year = 2017,
        month = jan,
       volume = {597},
          eid = {A131},
        pages = {A131},
          doi = {10.1051/0004-6361/201525659},
       adsurl = {https://ui.adsabs.harvard.edu/abs/2017A&A...597A.131W}
}

@article{planck_CS_2016,
   title={Planck2015 results},
   volume={594},
   ISSN={1432-0746},
   url={http://dx.doi.org/10.1051/0004-6361/201525967},
   DOI={10.1051/0004-6361/201525967},
   journal={Astronomy \& Astrophysics},
   publisher={EDP Sciences},
   author={{Planck results X}},
   year={2016},
   month={Sep},
   pages={A10}
}

@ARTICLE{mikes,
       author = {{Wilensky}, Michael J. and {Irfan}, Melis O. and {Bull}, Philip},
        title = "{Bayesian evidence for uncorrected gain factors in Galactic synchrotron template maps}",
      journal = {arXiv e-prints},
         year = 2024,
        month = sep,
          eid = {arXiv:2409.06770},
        pages = {arXiv:2409.06770},
          doi = {10.48550/arXiv.2409.06770},
archivePrefix = {arXiv},
       eprint = {2409.06770},
 primaryClass = {astro-ph.CO},
       adsurl = {https://ui.adsabs.harvard.edu/abs/2024arXiv240906770W}
}

@ARTICLE{TFpaper,
       author = {{Cunnington}, Steven and {Wolz}, Laura and {Bull}, Philip and {Carucci}, Isabella P. and {Grainge}, Keith and {Irfan}, Melis O. and {Li}, Yichao and {Pourtsidou}, Alkistis and {Santos}, Mario G. and {Spinelli}, Marta and {Wang}, Jingying},
        title = "{The foreground transfer function for H I intensity mapping signal reconstruction: MeerKLASS and precision cosmology applications}",
      journal = {MNRAS},
         year = 2023,
        month = aug,
       volume = {523},
       number = {2},
        pages = {2453-2477},
          doi = {10.1093/mnras/stad1567},
archivePrefix = {arXiv},
       eprint = {2302.07034},
 primaryClass = {astro-ph.CO},
       adsurl = {https://ui.adsabs.harvard.edu/abs/2023MNRAS.523.2453C}
}

@ARTICLE{mPCA,
    author = "Carucci, Isabella P. and others",
    title = "{Hydrogen intensity mapping with MeerKAT: Preserving cosmological signal by optimising contaminant separation}",
    eprint = "2412.06750",
    archivePrefix = "arXiv",
    primaryClass = "astro-ph.CO",
    doi = "10.1051/0004-6361/202453461",
    journal = "Astron. Astrophys.",
    volume = "703",
    pages = "A222",
    year = "2025"
}

@ARTICLE{2019MNRAS.489..385T,
       author = {{Tramonte}, Denis and {Ma}, Yin-Zhe and {Li}, Yi-Chao and {Staveley-Smith}, Lister},
        title = "{Searching for H I imprints in cosmic web filaments with 21-cm intensity mapping}",
      journal = {MNRAS},
         year = 2019,
        month = oct,
       volume = {489},
       number = {1},
        pages = {385-400},
          doi = {10.1093/mnras/stz2146},
archivePrefix = {arXiv},
       eprint = {1908.00028},
 primaryClass = {astro-ph.CO},
       adsurl = {https://ui.adsabs.harvard.edu/abs/2019MNRAS.489..385T}
}

@ARTICLE{2023ApJ...947...16A,
       author = {{CHIME Collaboration} and {Amiri}, Mandana and {Bandura}, Kevin and {Chen}, Tianyue and {Deng}, Meiling and {Dobbs}, Matt and {Fandino}, Mateus and {Foreman}, Simon and {Halpern}, Mark and {Hill}, Alex S. and {Hinshaw}, Gary and {H{\"o}fer}, Carolin and {Kania}, Joseph and {Landecker}, T.~L. and {MacEachern}, Joshua and {Masui}, Kiyoshi and {Mena-Parra}, Juan and {Milutinovic}, Nikola and {Mirhosseini}, Arash and {Newburgh}, Laura and {Ordog}, Anna and {Pen}, Ue-Li and {Pinsonneault-Marotte}, Tristan and {Polzin}, Ava and {Reda}, Alex and {Renard}, Andre and {Shaw}, J. Richard and {Siegel}, Seth R. and {Singh}, Saurabh and {Vanderlinde}, Keith and {Wang}, Haochen and {Wiebe}, Donald V. and {Wulf}, Dallas and {CHIME Collaboration}},
        title = "{Detection of Cosmological 21 cm Emission with the Canadian Hydrogen Intensity Mapping Experiment}",
      journal = {APJ},
         year = 2023,
        month = apr,
       volume = {947},
       number = {1},
          eid = {16},
        pages = {16},
          doi = {10.3847/1538-4357/acb13f},
archivePrefix = {arXiv},
       eprint = {2202.01242},
 primaryClass = {astro-ph.CO},
       adsurl = {https://ui.adsabs.harvard.edu/abs/2023ApJ...947...16A}
}

@ARTICLE{2025arXiv250403908C,
       author = {{Chen}, Zhaoting and {Cunnington}, Steven and {Pourtsidou}, Alkistis and {Wolz}, Laura and {Spinelli}, Marta and {Bernal}, Jos{\'e} Luis and {Barberi-Squarotti}, Matilde and {Camera}, Stefano and {Carucci}, Isabella P. and {Fonseca}, Jos{\'e} and {Grainge}, Keith and {Irfan}, Melis O. and {Santos}, Mario G. and {Wang}, Jingying},
        title = "{Emission line stacking of 21cm intensity maps with MeerKLASS: Inference pipeline and application to the L-band deep-field data}",
      journal = {arXiv e-prints},
         year = 2025,
        month = apr,
          eid = {arXiv:2504.03908},
        pages = {arXiv:2504.03908},
          doi = {10.48550/arXiv.2504.03908},
archivePrefix = {arXiv},
       eprint = {2504.03908},
 primaryClass = {astro-ph.CO},
       adsurl = {https://ui.adsabs.harvard.edu/abs/2025arXiv250403908C}
}

@ARTICLE{2025arXiv250321743D,
       author = {{Dunne}, D.~A. and {Cleary}, K.~A. and {Breysse}, P.~C. and {Chung}, D.~T. and {Ihle}, H.~T. and {Lunde}, J.~G.~S. and {Padmanabhan}, H. and {Stutzer}, N. -O. and {Bond}, J.~R. and {Gundersen}, J.~O. and {Kim}, J. and {Readhead}, A.~C.~S.},
        title = "{Three-Dimensional Stacking as a Line Intensity Mapping Statistic}",
      journal = {arXiv e-prints},
         year = 2025,
        month = mar,
          eid = {arXiv:2503.21743},
        pages = {arXiv:2503.21743},
          doi = {10.48550/arXiv.2503.21743},
archivePrefix = {arXiv},
       eprint = {2503.21743},
 primaryClass = {astro-ph.CO},
       adsurl = {https://ui.adsabs.harvard.edu/abs/2025arXiv250321743D}
}

@ARTICLE{2022MNRAS.514.4205S,
       author = {{Sinigaglia}, Francesco and {Elson}, Ed and {Rodighiero}, Giulia and {Vaccari}, Mattia},
        title = "{Optimizing spectral stacking for 21-cm observations of galaxies: accuracy assessment and symmetrized stacking}",
      journal = {MNRAS},
         year = 2022,
        month = aug,
       volume = {514},
       number = {3},
        pages = {4205-4221},
          doi = {10.1093/mnras/stac1584},
archivePrefix = {arXiv},
       eprint = {2206.03300},
 primaryClass = {astro-ph.CO},
       adsurl = {https://ui.adsabs.harvard.edu/abs/2022MNRAS.514.4205S}
}

@ARTICLE{2020MNRAS.498.5916T,
       author = {{Tramonte}, Denis and {Ma}, Yin-Zhe},
        title = "{The neutral hydrogen distribution in large-scale haloes from 21-cm intensity maps}",
      journal = {MNRAS},
         year = 2020,
        month = nov,
       volume = {498},
       number = {4},
        pages = {5916-5935},
          doi = {10.1093/mnras/staa2727},
archivePrefix = {arXiv},
       eprint = {2009.02387},
 primaryClass = {astro-ph.CO},
       adsurl = {https://ui.adsabs.harvard.edu/abs/2020MNRAS.498.5916T}
}

@ARTICLE{2024ApJ...965....7D,
       author = {{Dunne}, Delaney A. and {Cleary}, Kieran A. and {Breysse}, Patrick C. and {Chung}, Dongwoo T. and {Ihle}, H{\r{a}}vard T. and {Bond}, J. Richard and {Eriksen}, Hans Kristian and {Gundersen}, Joshua Ott and {Keating}, Laura C. and {Kim}, Junhan and {Lunde}, Jonas Gahr Sturtzel and {Murray}, Norman and {Padmanabhan}, Hamsa and {Philip}, Liju and {Stutzer}, Nils-Ole and {Tolgay}, Do{\u{g}}a and {Wehus}, Ingunn Katherine and {Church}, Sarah E. and {Gaier}, Todd and {Harris}, Andrew I. and {Hobbs}, Richard and {Lamb}, James W. and {Lawrence}, Charles R. and {Readhead}, Anthony C.~S. and {Woody}, David P.},
        title = "{COMAP Early Science. VIII. A Joint Stacking Analysis with eBOSS Quasars}",
      journal = {APJ},
         year = 2024,
        month = apr,
       volume = {965},
       number = {1},
          eid = {7},
        pages = {7},
          doi = {10.3847/1538-4357/ad2dfc},
archivePrefix = {arXiv},
       eprint = {2304.09832},
 primaryClass = {astro-ph.GA},
       adsurl = {https://ui.adsabs.harvard.edu/abs/2024ApJ...965....7D}
}

@ARTICLE{2022ApJ...929...90L,
       author = {{Lujan Niemeyer}, Maja and {Komatsu}, Eiichiro and {Byrohl}, Chris and {Davis}, Dustin and {Fabricius}, Maximilian and {Gebhardt}, Karl and {Hill}, Gary J. and {Wisotzki}, Lutz and {Bowman}, William P. and {Ciardullo}, Robin and {Farrow}, Daniel J. and {Finkelstein}, Steven L. and {Gawiser}, Eric and {Gronwall}, Caryl and {Jeong}, Donghui and {Landriau}, Martin and {Liu}, Chenxu and {Cooper}, Erin Mentuch and {Ouchi}, Masami and {Schneider}, Donald P. and {Zeimann}, Gregory R.},
        title = "{Surface Brightness Profile of Lyman-{\ensuremath{\alpha}} Halos out to 320 kpc in HETDEX}",
      journal = {APJ},
         year = 2022,
        month = apr,
       volume = {929},
       number = {1},
          eid = {90},
        pages = {90},
          doi = {10.3847/1538-4357/ac5cb8},
archivePrefix = {arXiv},
       eprint = {2203.04826},
 primaryClass = {astro-ph.GA},
       adsurl = {https://ui.adsabs.harvard.edu/abs/2022ApJ...929...90L}
}

@ARTICLE{2022ApJ...934L..26L,
       author = {{Lujan Niemeyer}, Maja and {Bowman}, William P. and {Ciardullo}, Robin and {Gronke}, Max and {Komatsu}, Eiichiro and {Fabricius}, Maximilian and {Farrow}, Daniel J. and {Finkelstein}, Steven L. and {Gebhardt}, Karl and {Gronwall}, Caryl and {Hill}, Gary J. and {Liu}, Chenxu and {Cooper}, Erin Mentuch and {Schneider}, Donald P. and {Tuttle}, Sarah and {Zeimann}, Gregory R.},
        title = "{Ly{\ensuremath{\alpha}} Halos around [O III]-selected Galaxies in HETDEX}",
      journal = {APJl},
         year = 2022,
        month = aug,
       volume = {934},
       number = {2},
          eid = {L26},
        pages = {L26},
          doi = {10.3847/2041-8213/ac82e5},
archivePrefix = {arXiv},
       eprint = {2207.11098},
 primaryClass = {astro-ph.GA},
       adsurl = {https://ui.adsabs.harvard.edu/abs/2022ApJ...934L..26L}
}

@ARTICLE{6410347,
  author={de Villiers, Dirk I. L.},
  journal={IEEE Transactions on Antennas and Propagation}, 
  title={Prediction of Aperture Efficiency Ripple in Clear Aperture Offset Gregorian Antennas}, 
  year={2013},
  volume={61},
  number={5},
  pages={2457-2465},
  doi={10.1109/TAP.2013.2239953}}

@ARTICLE{2021MNRAS.502.2970A,
       author = {{Asad}, K.~M.~B. and {Girard}, J.~N. and {de Villiers}, M. and {Ansah-Narh}, T. and {Iheanetu}, K. and {Smirnov}, O. and {Santos}, M.~G. and {Lehmensiek}, R. and {Jonas}, J. and {de Villiers}, D.~I.~L. and {Thorat}, K. and {Hugo}, B. and {Makhathini}, S. and {Jozsa}, G.~I.~G. and {Sirothia}, S.~K.},
        title = "{Primary beam effects of radio astronomy antennas - II. Modelling MeerKAT L-band beams}",
      journal = {MNRAS},
         year = 2021,
        month = apr,
       volume = {502},
       number = {2},
        pages = {2970-2983},
          doi = {10.1093/mnras/stab104},
archivePrefix = {arXiv},
       eprint = {1904.07155},
 primaryClass = {astro-ph.IM},
       adsurl = {https://ui.adsabs.harvard.edu/abs/2021MNRAS.502.2970A}
}

@ARTICLE{2004MNRAS.347..645P,
       author = {{Percival}, Will J. and {Verde}, Licia and {Peacock}, John A.},
        title = "{Fourier analysis of luminosity-dependent galaxy clustering}",
      journal = {MNRAS},
         year = 2004,
        month = jan,
       volume = {347},
       number = {2},
        pages = {645-653},
          doi = {10.1111/j.1365-2966.2004.07245.x},
archivePrefix = {arXiv},
       eprint = {astro-ph/0306511},
 primaryClass = {astro-ph},
       adsurl = {https://ui.adsabs.harvard.edu/abs/2004MNRAS.347..645P}
}

@ARTICLE{2009MNRAS.397.1348W,
       author = {{White}, Martin and {Song}, Yong-Seon and {Percival}, Will J.},
        title = "{Forecasting cosmological constraints from redshift surveys}",
      journal = {MNRAS},
         year = 2009,
        month = aug,
       volume = {397},
       number = {3},
        pages = {1348-1354},
          doi = {10.1111/j.1365-2966.2008.14379.x},
archivePrefix = {arXiv},
       eprint = {0810.1518},
 primaryClass = {astro-ph},
       adsurl = {https://ui.adsabs.harvard.edu/abs/2009MNRAS.397.1348W}
}

@ARTICLE{2009PhRvL.102b1302S,
       author = {{Seljak}, Uro{\v{s}}},
        title = "{Extracting Primordial Non-Gaussianity without Cosmic Variance}",
      journal = {Physical Review Letters},
         year = 2009,
        month = jan,
       volume = {102},
       number = {2},
          eid = {021302},
        pages = {021302},
          doi = {10.1103/PhysRevLett.102.021302},
archivePrefix = {arXiv},
       eprint = {0807.1770},
 primaryClass = {astro-ph},
       adsurl = {https://ui.adsabs.harvard.edu/abs/2009PhRvL.102b1302S}
}

@ARTICLE{2009JCAP...10..007M,
       author = {{McDonald}, Patrick and {Seljak}, Uro{\v{s}}},
        title = "{How to evade the sample variance limit on measurements of redshift-space distortions}",
      journal = {JCAP},
         year = 2009,
        month = oct,
       volume = {2009},
       number = {10},
          eid = {007},
        pages = {007},
          doi = {10.1088/1475-7516/2009/10/007},
archivePrefix = {arXiv},
       eprint = {0810.0323},
 primaryClass = {astro-ph},
       adsurl = {https://ui.adsabs.harvard.edu/abs/2009JCAP...10..007M}
}

@article{Bharadwaj:2000av,
    author = "Bharadwaj, Somnath and Nath, B.B. and Nath, Biman B. and Sethi, Shiv K.",
    title = "{Using HI to probe large scale structures at z \textasciitilde\ 3}",
    eprint = "astro-ph/0003200",
    archivePrefix = "arXiv",
    doi = "10.1007/BF02933588",
    journal = "J. Astrophys. Astron.",
    volume = "22",
    pages = "21",
    year = "2001"
}

@article{Battye:2004re,
    author = "Battye, Richard A. and Davies, Rod D. and Weller, Jochen",
    title = "{Neutral hydrogen surveys for high redshift galaxy clusters and proto-clusters}",
    eprint = "astro-ph/0401340",
    archivePrefix = "arXiv",
    doi = "10.1111/j.1365-2966.2004.08416.x",
        journal = "MNRAS",
    volume = "355",
    pages = "1339--1347",
    year = "2004"
}

@article{Chang:2007xk,
    author = "Chang, Tzu-Ching and Pen, Ue-Li and Peterson, Jeffrey B. and McDonald, Patrick",
    title = "{Baryon Acoustic Oscillation Intensity Mapping as a Test of Dark Energy}",
    eprint = "0709.3672",
    archivePrefix = "arXiv",
    primaryClass = "astro-ph",
    doi = "10.1103/PhysRevLett.100.091303",
    journal = "Phys. Rev. Lett.",
    volume = "100",
    pages = "091303",
    year = "2008"
}

@article{Wyithe:2007rq,
    author = "Wyithe, Stuart and Loeb, Abraham and Geil, Paul",
    title = "{Baryonic Acoustic Oscillations in 21cm Emission: A Probe of Dark Energy out to High Redshifts}",
    eprint = "0709.2955",
    archivePrefix = "arXiv",
    primaryClass = "astro-ph",
    doi = "10.1111/j.1365-2966.2007.12631.x",
    journal = "MNRAS",
    volume = "383",
    pages = "1195",
    year = "2008"
}

@article{Liu:2019awk,
    author = "Liu, Adrian and Shaw, J. Richard",
    title = "{Data Analysis for Precision 21 cm Cosmology}",
    eprint = "1907.08211",
    archivePrefix = "arXiv",
    primaryClass = "astro-ph.IM",
    doi = "10.1088/1538-3873/ab5bfd",
    journal = "Publ. Astron. Soc. Pac.",
    volume = "132",
    number = "1012",
    pages = "062001",
    year = "2020"
}

@article{Villaescusa-Navarro:2018vsg,
    author = "Villaescusa-Navarro, Francisco and others",
    title = "{Ingredients for 21 cm Intensity Mapping}",
    eprint = "1804.09180",
    archivePrefix = "arXiv",
    primaryClass = "astro-ph.CO",
    doi = "10.3847/1538-4357/aadba0",
    journal = "Astrophys. J.",
    volume = "866",
    number = "2",
    pages = "135",
    year = "2018"
}

@article{Kovetz:2017agg,
    author = "Kovetz, Ely D. and others",
    title = "{Line-Intensity Mapping: 2017 Status Report}",
    eprint = "1709.09066",
    archivePrefix = "arXiv",
    primaryClass = "astro-ph.CO",
    month = "9",
    year = "2017"
}

@article{Zaroubi:2012in,
    author = "Zaroubi, Saleem",
    title = "{The Epoch of Reionization}",
    eprint = "1206.0267",
    archivePrefix = "arXiv",
    primaryClass = "astro-ph.CO",
    doi = "10.1007/978-3-642-32362-1_2",
    month = "6",
    year = "2012"
}

@article{Furlanetto:2006jb,
    author = "Furlanetto, Steven and Oh, S. Peng and Briggs, Frank",
    title = "{Cosmology at Low Frequencies: The 21 cm Transition and the High-Redshift Universe}",
    eprint = "astro-ph/0608032",
    archivePrefix = "arXiv",
    doi = "10.1016/j.physrep.2006.08.002",
    journal = "Phys. Rept.",
    volume = "433",
    pages = "181--301",
    year = "2006"
}

@article{Scoccimarro:2004tg,
    author = "Scoccimarro, Roman",
    title = "{Redshift-space distortions, pairwise velocities and nonlinearities}",
    eprint = "astro-ph/0407214",
    archivePrefix = "arXiv",
    doi = "10.1103/PhysRevD.70.083007",
    journal = "Phys. Rev. D",
    volume = "70",
    pages = "083007",
    year = "2004"
}

@article{Eisenstein:1997ik,
    author = "Eisenstein, Daniel J. and Hu, Wayne",
    title = "{Baryonic features in the matter transfer function}",
    eprint = "astro-ph/9709112",
    archivePrefix = "arXiv",
    reportNumber = "IASSNS-AST-97-51",
    doi = "10.1086/305424",
    journal = "Astrophys. J.",
    volume = "496",
    pages = "605",
    year = "1998"
}

@article{Koyama:2015vza,
    author = "Koyama, Kazuya",
    title = "{Cosmological Tests of Modified Gravity}",
    eprint = "1504.04623",
    archivePrefix = "arXiv",
    primaryClass = "astro-ph.CO",
    doi = "10.1088/0034-4885/79/4/046902",
    journal = "Rept. Prog. Phys.",
    volume = "79",
    number = "4",
    pages = "046902",
    year = "2016"
}

@article{Dalal:2007cu,
    author = "Dalal, Neal and Dore, Olivier and Huterer, Dragan and Shirokov, Alexander",
    title = "{The imprints of primordial non-gaussianities on large-scale structure: scale dependent bias and abundance of virialized objects}",
    eprint = "0710.4560",
    archivePrefix = "arXiv",
    primaryClass = "astro-ph",
    doi = "10.1103/PhysRevD.77.123514",
    journal = "Phys. Rev. D",
    volume = "77",
    pages = "123514",
    year = "2008"
}

@article{Bonvin:2011bg,
    author = "Bonvin, Camille and Durrer, Ruth",
    title = "{What galaxy surveys really measure}",
    eprint = "1105.5280",
    archivePrefix = "arXiv",
    primaryClass = "astro-ph.CO",
    doi = "10.1103/PhysRevD.84.063505",
    journal = "Phys. Rev. D",
    volume = "84",
    pages = "063505",
    year = "2011"
}

@article{Bull:2014rha,
    author = "Bull, Philip and Ferreira, Pedro G. and Patel, Prina and Santos, Mario G.",
    title = "{Late-time cosmology with 21cm intensity mapping experiments}",
    eprint = "1405.1452",
    archivePrefix = "arXiv",
    primaryClass = "astro-ph.CO",
    doi = "10.1088/0004-637X/803/1/21",
    journal = "Astrophys. J.",
    volume = "803",
    number = "1",
    pages = "21",
    year = "2015"
}

@INPROCEEDINGS{2015aska.confE..19S,
       author = {{Santos}, M. and {Bull}, P. and {Alonso}, D. and {Camera}, S. and {Ferreira}, P. and {Bernardi}, G. and {Maartens}, R. and {Viel}, M. and {Villaescusa-Navarro}, F. and {Abdalla}, F.~B. and {Jarvis}, M. and {Metcalf}, R.~B. and {Pourtsidou}, A. and {Wolz}, L.},
        title = "{Cosmology from a SKA HI intensity mapping survey}",
    booktitle = {Advancing Astrophysics with the Square Kilometre Array (AASKA14)},
         year = 2015,
        month = apr,
          eid = {19},
        pages = {19},
          doi = {10.22323/1.215.0019},
archivePrefix = {arXiv},
       eprint = {1501.03989},
 primaryClass = {astro-ph.CO},
       adsurl = {https://ui.adsabs.harvard.edu/abs/2015aska.confE..19S}
}

@inproceedings{MeerKLASS:2017vgf,
    author = "Santos, Mario G. and others",
    collaboration = "MeerKLASS",
    title = "{MeerKLASS: MeerKAT Large Area Synoptic Survey}",
    booktitle = "MeerKAT Science: On the Pathway to the SKA",
    eprint = "1709.06099",
    archivePrefix = "arXiv",
    primaryClass = "astro-ph.CO",
    month = "9",
    year = "2017"
}

@article{MatildeAutoPk,
    author = "{MeerKLASS Collaboration} and others",
    title = "{HI intensity mapping with MeerKAT: An auto-power spectrum measurement consistent with a cosmological 21-cm signal}",
    year = "in prep."
}

@article{Sinigaglia:2025weq,
    author = "Sinigaglia, Francesco and Bianchetti, Alessandro and Rodighiero, Giulia and Mayer, Lucio and Dessauges-Zavadsky, Miroslava and Elson, E. and Vaccari, Mattia and Jarvis, Matt J.",
    title = "{Semiempirical constraints on the HI mass function of star-forming galaxies and {\ensuremath{\Omega}}HI at z{\ensuremath{\sim}} 0.37 from interferometric surveys}",
    eprint = "2506.11280",
    archivePrefix = "arXiv",
    primaryClass = "astro-ph.GA",
    doi = "10.1051/0004-6361/202555928",
    journal = "Astron. Astrophys.",
    volume = "704",
    pages = "A152",
    year = "2025"
}

@article{Paul:2023yrr,
    author = "Paul, Sourabh and Santos, Mario G. and Chen, Zhaoting and Wolz, Laura",
    title = "{A first detection of neutral hydrogen intensity mapping on Mpc scales at $z\approx 0.32$ and $z\approx 0.44$}",
    eprint = "2301.11943",
    archivePrefix = "arXiv",
    primaryClass = "astro-ph.CO",
    month = "1",
    year = "2023"
}

@article{DESI:2016fyo,
    author = "{DESI Collaboration} and others",
    collaboration = "DESI",
    title = "{The DESI Experiment Part I: Science,Targeting, and Survey Design}",
    eprint = "1611.00036",
    archivePrefix = "arXiv",
    primaryClass = "astro-ph.IM",
    reportNumber = "FERMILAB-PUB-16-517-AE",
    month = "10",
    year = "2016"
}

@ARTICLE{2019Msngr.175....3D,
       author = {{4MOST Collaboration} and others},
        title = "{4MOST: Project overview and information for the First Call for Proposals}",
      journal = {The Messenger},
         year = 2019,
        month = mar,
       volume = {175},
        pages = {3-11},
          doi = {10.18727/0722-6691/5117},
archivePrefix = {arXiv},
       eprint = {1903.02464},
 primaryClass = {astro-ph.IM},
       adsurl = {https://ui.adsabs.harvard.edu/abs/2019Msngr.175....3D}
}

@article{Euclid:2024yrr,
    author = "{Euclid Collaboration} and others",
    collaboration = "Euclid",
    title = "{Euclid. I. Overview of the Euclid mission}",
    eprint = "2405.13491",
    archivePrefix = "arXiv",
    primaryClass = "astro-ph.CO",
    doi = "10.1051/0004-6361/202450810",
    journal = "Astron. Astrophys.",
    volume = "697",
    pages = "A1",
    year = "2025"
}

@article{LSSTDarkEnergyScience:2018jkl,
    author = "{LSST DESC} and others",
    collaboration = "LSST Dark Energy Science",
    title = "{The LSST Dark Energy Science Collaboration (DESC) Science Requirements Document}",
    eprint = "1809.01669",
    archivePrefix = "arXiv",
    primaryClass = "astro-ph.CO",
    reportNumber = "FERMILAB-PUB-18-465-A",
    doi = "10.2172/1471560",
    month = "9",
    year = "2018"
}

@article{SKA:2018ckk,
    author = "{SKAO Cosmology SWG} and others",
    collaboration = "SKA",
    title = "{Cosmology with Phase 1 of the Square Kilometre Array: Red Book 2018: Technical specifications and performance forecasts}",
    eprint = "1811.02743",
    archivePrefix = "arXiv",
    primaryClass = "astro-ph.CO",
    doi = "10.1017/pasa.2019.51",
    journal = "Publ. Astron. Soc. Austral.",
    volume = "37",
    pages = "e007",
    year = "2020"
}

@INPROCEEDINGS{8738133,
  author={Lehmensiek, Robert and de Villiers, Dirk I.L.},
  booktitle={2019 URSI Asia-Pacific Radio Science Conference (AP-RASC)}, 
  title={On the performance of the SKA mid-frequency array’s reflector system and its feeds}, 
  year={2019},
  volume={},
  number={},
  pages={1-3},
  doi={10.23919/URSIAP-RASC.2019.8738133}}

@article{Masui:2012zc,
    author = "Masui, K. W. and others",
    title = "{Measurement of 21 cm brightness fluctuations at z {\textasciitilde} 0.8 in cross-correlation}",
    eprint = "1208.0331",
    archivePrefix = "arXiv",
    primaryClass = "astro-ph.CO",
    doi = "10.1088/2041-8205/763/1/L20",
    journal = "Astrophys. J. Lett.",
    volume = "763",
    pages = "L20",
    year = "2013"
}

@article{Switzer:2013ewa,
    author = "Switzer, E. R. and others",
    title = "{Determination of z{\textasciitilde}0.8 neutral hydrogen fluctuations using the 21 cm intensity mapping auto-correlation}",
    eprint = "1304.3712",
    archivePrefix = "arXiv",
    primaryClass = "astro-ph.CO",
    doi = "10.1093/mnrasl/slt074",
    journal = "Mon. Not. Roy. Astron. Soc.",
    volume = "434",
    pages = "L46",
    year = "2013"
}

@article{Wolz:2015lwa,
    author = "Wolz, L. and others",
    title = "{Erasing the Milky Way: new cleaning technique applied to GBT intensity mapping data}",
    eprint = "1510.05453",
    archivePrefix = "arXiv",
    primaryClass = "astro-ph.CO",
    doi = "10.1093/mnras/stw2556",
    journal = "Mon. Not. Roy. Astron. Soc.",
    volume = "464",
    number = "4",
    pages = "4938--4949",
    year = "2017"
}

@article{Anderson:2017ert,
    author = "Anderson, C. J. and others",
    title = "{Low-amplitude clustering in low-redshift 21-cm intensity maps cross-correlated with 2dF galaxy densities}",
    eprint = "1710.00424",
    archivePrefix = "arXiv",
    primaryClass = "astro-ph.CO",
    doi = "10.1093/mnras/sty346",
    journal = "Mon. Not. Roy. Astron. Soc.",
    volume = "476",
    number = "3",
    pages = "3382--3392",
    year = "2018"
}

@article{eBOSS:2021ebm,
    author = "Wolz, Laura and others",
    collaboration = "eBOSS",
    title = "{H{\,}i constraints from the cross-correlation of eBOSS galaxies and Green Bank Telescope intensity maps}",
    eprint = "2102.04946",
    archivePrefix = "arXiv",
    primaryClass = "astro-ph.CO",
    doi = "10.1093/mnras/stab3621",
    journal = "Mon. Not. Roy. Astron. Soc.",
    volume = "510",
    number = "3",
    pages = "3495--3511",
    year = "2022"
}

@INPROCEEDINGS{2016mks..confE...1J,
       author = {{Jonas}, J. and {MeerKAT Team}},
        title = "{The MeerKAT Radio Telescope}",
    booktitle = {MeerKAT Science: On the Pathway to the SKA},
         year = 2016,
        month = jan,
          eid = {1},
        pages = {1},
          doi = {10.22323/1.277.0001},
       adsurl = {https://ui.adsabs.harvard.edu/abs/2016mks..confE...1J}
}

@article{CHIME:2022dwe,
    author = "Amiri, Mandana and others",
    collaboration = "CHIME",
    title = "{An Overview of CHIME, the Canadian Hydrogen Intensity Mapping Experiment}",
    eprint = "2201.07869",
    archivePrefix = "arXiv",
    primaryClass = "astro-ph.IM",
    doi = "10.3847/1538-4365/ac6fd9",
    journal = "Astrophys. J. Supp.",
    volume = "261",
    number = "2",
    pages = "29",
    year = "2022"
}

@article{CHIME:2022kvg,
    author = "Amiri, Mandana and others",
    collaboration = "CHIME",
    title = "{Detection of Cosmological 21 cm Emission with the Canadian Hydrogen Intensity Mapping Experiment}",
    eprint = "2202.01242",
    archivePrefix = "arXiv",
    primaryClass = "astro-ph.CO",
    doi = "10.3847/1538-4357/acb13f",
    journal = "Astrophys. J.",
    volume = "947",
    number = "1",
    pages = "16",
    year = "2023"
}

@article{CHIME:2023til,
    author = "Amiri, Mandana and others",
    collaboration = "CHIME",
    title = "{A Detection of Cosmological 21 cm Emission from CHIME in Cross-correlation with eBOSS Measurements of the Ly{\ensuremath{\alpha}} Forest}",
    eprint = "2309.04404",
    archivePrefix = "arXiv",
    primaryClass = "astro-ph.CO",
    doi = "10.3847/1538-4357/ad0f1d",
    journal = "Astrophys. J.",
    volume = "963",
    number = "1",
    pages = "23",
    year = "2024"
}

@article{CHIME:2025cee,
    author = "Amiri, Mandana and others",
    collaboration = "CHIME",
    title = "{Detection of the Cosmological 21 cm Signal in Auto-correlation at z {\textasciitilde} 1 with the Canadian Hydrogen Intensity Mapping Experiment}",
    eprint = "2511.19620",
    archivePrefix = "arXiv",
    primaryClass = "astro-ph.CO",
    journal = "arXiv:",
    month = "11",
    year = "2025"
}

@article{Elahi:2024qbp,
    author = "Elahi, Khandakar Md Asif and others",
    title = "{Towards 21-cm intensity mapping at z~= 2.28 with uGMRT using the tapered gridded estimator {\textendash} IV. Wide-band analysis}",
    eprint = "2403.06736",
    archivePrefix = "arXiv",
    primaryClass = "astro-ph.CO",
    doi = "10.1093/mnras/stae740",
    journal = "Mon. Not. Roy. Astron. Soc.",
    volume = "529",
    number = "4",
    pages = "3372--3386",
    year = "2024"
}

@ARTICLE{2020MNRAS.493.5854H,
       author = {{Hu}, Wenkai and {Wang}, Xin and {Wu}, Fengquan and {Wang}, Yougang and {Zhang}, Pengjie and {Chen}, Xuelei},
        title = "{Forecast for FAST: from galaxies survey to intensity mapping}",
    journal = "Mon. Not. Roy. Astron. Soc.",
         year = 2020,
        month = apr,
       volume = {493},
       number = {4},
        pages = {5854-5870},
          doi = {10.1093/mnras/staa650},
archivePrefix = {arXiv},
       eprint = {1909.10946},
 primaryClass = {astro-ph.CO},
       adsurl = {https://ui.adsabs.harvard.edu/abs/2020MNRAS.493.5854H}
}

@article{Li:2023zer,
    author = "Li, Yichao and others",
    title = "{FAST Drift Scan Survey for Hi Intensity Mapping: I. Preliminary Data Analysis}",
    eprint = "2305.06405",
    archivePrefix = "arXiv",
    primaryClass = "astro-ph.CO",
    reportNumber = "FERMILAB-PUB-23-258-T",
    doi = "10.3847/1538-4357/ace896",
    journal = "Astrophys. J.",
    volume = "954",
    number = "2",
    pages = "139",
    year = "2023"
}

@article{Mangla:2025ans,
    author = "Mangla, Sarvesh and others",
    title = "{The MeerKLASS L-band On-the-Fly Continuum Survey: Data Release 1}",
    eprint = "2512.17685",
    archivePrefix = "arXiv",
    primaryClass = "astro-ph.IM",
    month = "12",
    year = "2025"
}

@article{Paul:2025cjp,
    author = "Paul, Sourabh and others",
    title = "{The MeerKLASS UHF On-the-Fly Continuum Survey -- Data Release I}",
    eprint = "2512.11964",
    archivePrefix = "arXiv",
    primaryClass = "astro-ph.GA",
    month = "12",
    year = "2025"
}

@article{Chatterjee:2025wzm,
    author = "Chatterjee, Suman and others",
    title = "{The MeerKLASS On-the-Fly continuum survey: pipeline design and validation}",
    eprint = "2512.11978",
    archivePrefix = "arXiv",
    primaryClass = "astro-ph.IM",
    month = "12",
    year = "2025"
}

@article{Cunnington:2025sdr,
    author = "Cunnington, Steven and others",
    title = "{Revealing cosmological fluctuations in 21 cm intensity maps with MeerKLASS: from maps to power spectra}",
    eprint = "2510.27549",
    archivePrefix = "arXiv",
    primaryClass = "astro-ph.CO",
    doi = "10.1007/s10509-026-04547-7",
    journal = "Astrophys. Space Sci.",
    volume = "371",
    number = "2",
    pages = "16",
    year = "2026"
}

@BOOK{2013tra..book.....W,
       author = {{Wilson}, Thomas L. and {Rohlfs}, Kristen and {H{\"u}ttemeister}, Susanne},
        title = "{Tools of Radio Astronomy}",
         year = 2013,
          doi = {10.1007/978-3-642-39950-3},
       adsurl = {https://ui.adsabs.harvard.edu/abs/2013tra..book.....W}
}

@incollection{Fonseca01.2026.SKA, author = {José Fonseca and author2 and author3 and author4 and author5},title = {},year = {2026},publisher = {},note = {arXiv search: Report number AASKAII/Fonseca01},booktitle = {Advancing Astrophysics with the SKA -- II (AASKAII)}}

@incollection{Elahi01.2026.SKA, author = {Khandakar Md Asif Elahi and author2 and author3 and author4 and author5},title = {},year = {2026},publisher = {},note = {arXiv search: Report number AASKAII/Elahi01},booktitle = {Advancing Astrophysics with the SKA -- II (AASKAII)}}

@incollection{Spinelli01.2026.SKA, author = {Marta Spinelli and author2 and author3 and author4 and author5},title = {},year = {2026},publisher = {},note = {arXiv search: Report number AASKAII/Spinelli01},booktitle = {Advancing Astrophysics with the SKA -- II (AASKAII)}}

@incollection{Chatterjee01.2026.SKA, author = {Suman Chatterjee and author2 and author3 and author4 and author5},title = {},year = {2026},publisher = {},note = {arXiv search: Report number AASKAII/Chatterjee01},booktitle = {Advancing Astrophysics with the SKA -- II (AASKAII)}}

@incollection{Wolz01.2026.SKA, author = {Laura Wolz and author2 and author3 and author4 and author5},title = {},year = {2026},publisher = {},note = {arXiv search: Report number AASKAII/Wolz01},booktitle = {Advancing Astrophysics with the SKA -- II (AASKAII)}}

\end{document}